\documentclass[a4paper,twocolumn,superscriptaddress,11pt,accepted=2020-09-29]{quantumarticle}
\pdfoutput=1
\usepackage[utf8]{inputenc}
\usepackage[english]{babel}
\usepackage[T1]{fontenc}
\usepackage{tabularx}
\usepackage{amsmath}
\usepackage{hyperref}
\usepackage{caption}
\usepackage{standalone}
\usepackage{tikz}
\usepackage{lipsum}
\usepackage{jlcode}
\usepackage{qcircuit}
\usepackage{upquote}
\usepackage[numbers,sort&compress]{natbib}
\usepackage{soul}
\usepackage{listings}
\usepackage{subcaption}
\usepackage{graphicx}
\usetikzlibrary{arrows,shapes,chains,trees}
\usepackage[T1]{fontenc}
\usepackage[normalem]{ulem}
\usepackage{listings}
\usepackage{booktabs}
\usepackage[framemethod=TikZ]{mdframed}

\newcommand{\Eq}[1]{Eq.~(\ref{#1})}
\newcommand{\Fig}[1]{Fig.~\ref{#1}}
\newcommand{\Lst}[1]{Listing~\ref{#1}}
\newcommand{\App}[1]{Appendix~\ref{#1}}
\newcommand{\Tbl}[1]{Table~\ref{#1}}

\newcommand{\expect}[2]{{\mathop{\mathbb{E}}\limits_{\substack{#2}}\left[#1\right]}}

\newcommand{\ra}[1]{\renewcommand{\arraystretch}{#1}}

\begin{document}

\title{Yao.jl: Extensible, Efficient Framework for Quantum Algorithm Design}
\date{\today}
\author[1, 2, 3, 4]{Xiu-Zhe Luo}
\email{rogerluo.rl18@gmail.com}
\author[1]{Jin-Guo Liu}
\email{cacate0129@iphy.ac.cn}
\author[2]{Pan Zhang}
\author[1, 5]{Lei Wang}

\affil[1]{Institute of Physics, Chinese Academy of Sciences, Beijing 100190, China}
\affil[2]{Institute of Theoretical Physics, Chinese Academy of Sciences, Beijing 100190, China}
\affil[3]{Department of Physics and Astronomy, University of Waterloo, Waterloo N2L 3G1, Canada}
\affil[4]{Perimeter Institute for Theoretical Physics, Waterloo, Ontario N2L 2Y5, Canada}
\affil[5]{Songshan Lake Materials Laboratory, Dongguan, Guangdong 523808, China}

\maketitle

\begin{abstract}
We introduce \texttt{Yao}, an extensible, efficient open-source framework for
quantum algorithm design. \texttt{Yao} features generic and differentiable programming of quantum circuits. It achieves state-of-the-art performance in simulating small to intermediate-sized quantum circuits that are relevant to near-term applications. We introduce the design principles and critical techniques behind \texttt{Yao}. These include the quantum block intermediate representation of quantum circuits, a builtin automatic differentiation engine optimized for reversible computing, and batched quantum registers with GPU acceleration. The extensibility and efficiency of \texttt{Yao} help boost innovation in quantum algorithm design. 
\end{abstract}

\section{Introduction}
\begin{figure}[t]
    \centerline{\includegraphics[width=3cm]{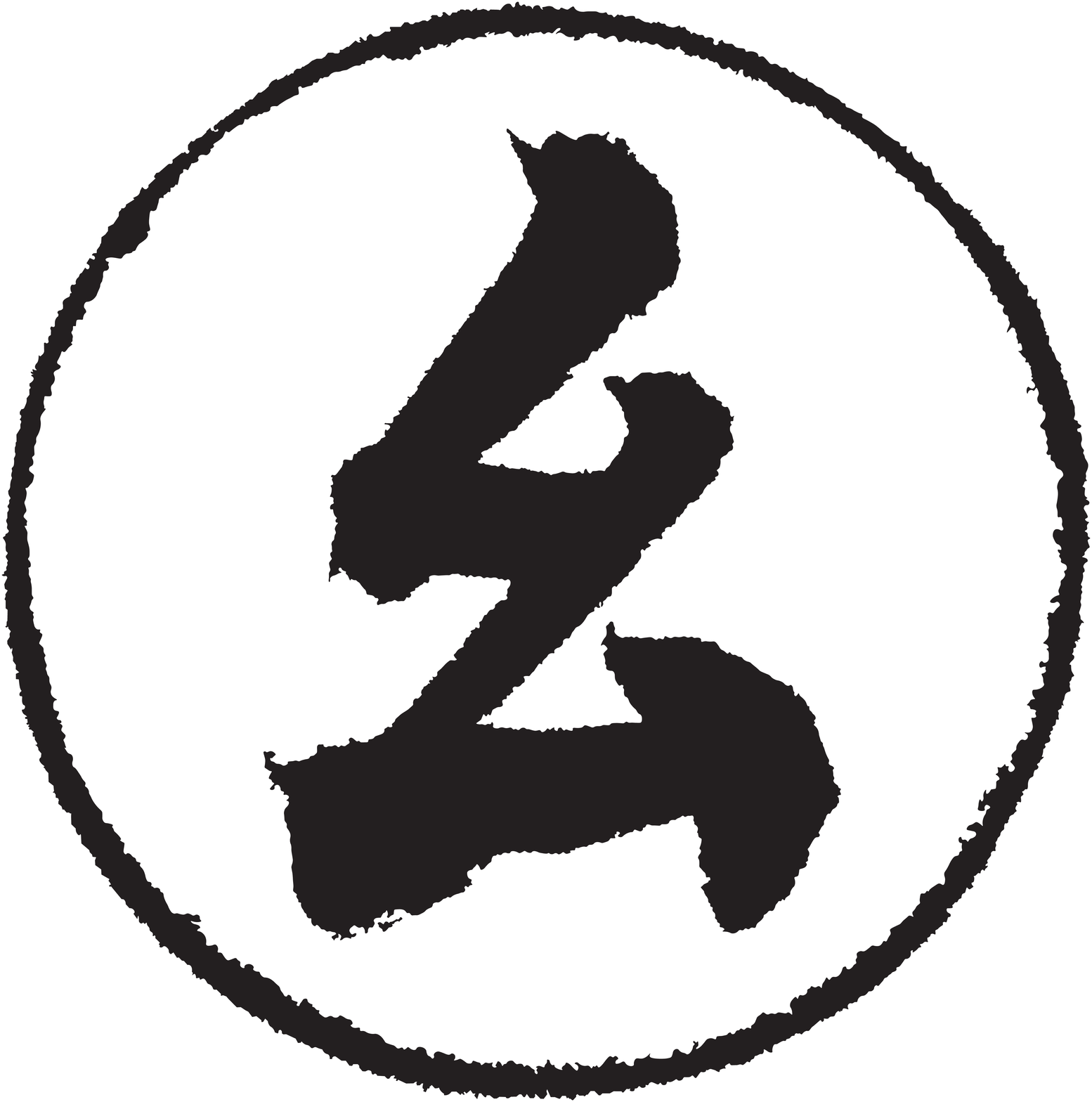}}
    \caption{
The logo of \texttt{Yao}. It contains a Chinese character pronounced as \texttt{y\=ao}, which stands for being unitary.
    }
    \label{fig:logo}
\end{figure}

\texttt{Yao} is a software for solving practical problems in quantum computation research. Given the limitations of near-term noisy intermediate-scale quantum circuits~\cite{preskill2018quantum}, it is advantageous to treat quantum devices as co-processors and compliment their abilities with classical computing resources.
Variational quantum algorithms have emerged as a promising research direction in particular.
These algorithms typically involve a quantum circuit with adjustable gate parameters and a classical optimizer. 
Many of these quantum algorithms, including the variational quantum eigensolver for ground states~\cite{peruzzo2014variational, wecker2015progress, mcclean2016theory}, quantum approximate optimization
algorithm for combinatorial problems~\cite{farhi2014quantum}, quantum circuit learning for classification and regression~\cite{2018arXiv180206002F, mitarai2018quantum}, 
and quantum circuit Born machine for generative modeling~\cite{Benedetti2019,liu2018differentiable} have had small scale demonstrations in experiments~\cite{o2016scalable, Kandala2017, havlivcek2019supervised, zhu2019training, qaoa-exp19, leyton2019robust}.
There are still fundamental issues in this field that call for better quantum software alongside hardware advances. 
For example, variational optimization of random circuits may encounter exponentially vanishing gradients~\cite{McClean2018} as the qubit number increases. Efficient quantum software is crucial for designing and verifying quantum algorithms in these challenging regimes.
Other research demands also call for quantum software that features a small overhead for repeated feedback control, convenient circuit structure manipulations, and efficient gradient calculation besides simply pushing up the number of qubits in experiments. 

\begin{figure*}[t]
    \centering
    \centerline{\includegraphics[width=\textwidth, trim={0 5cm 2cm 1cm}, clip]{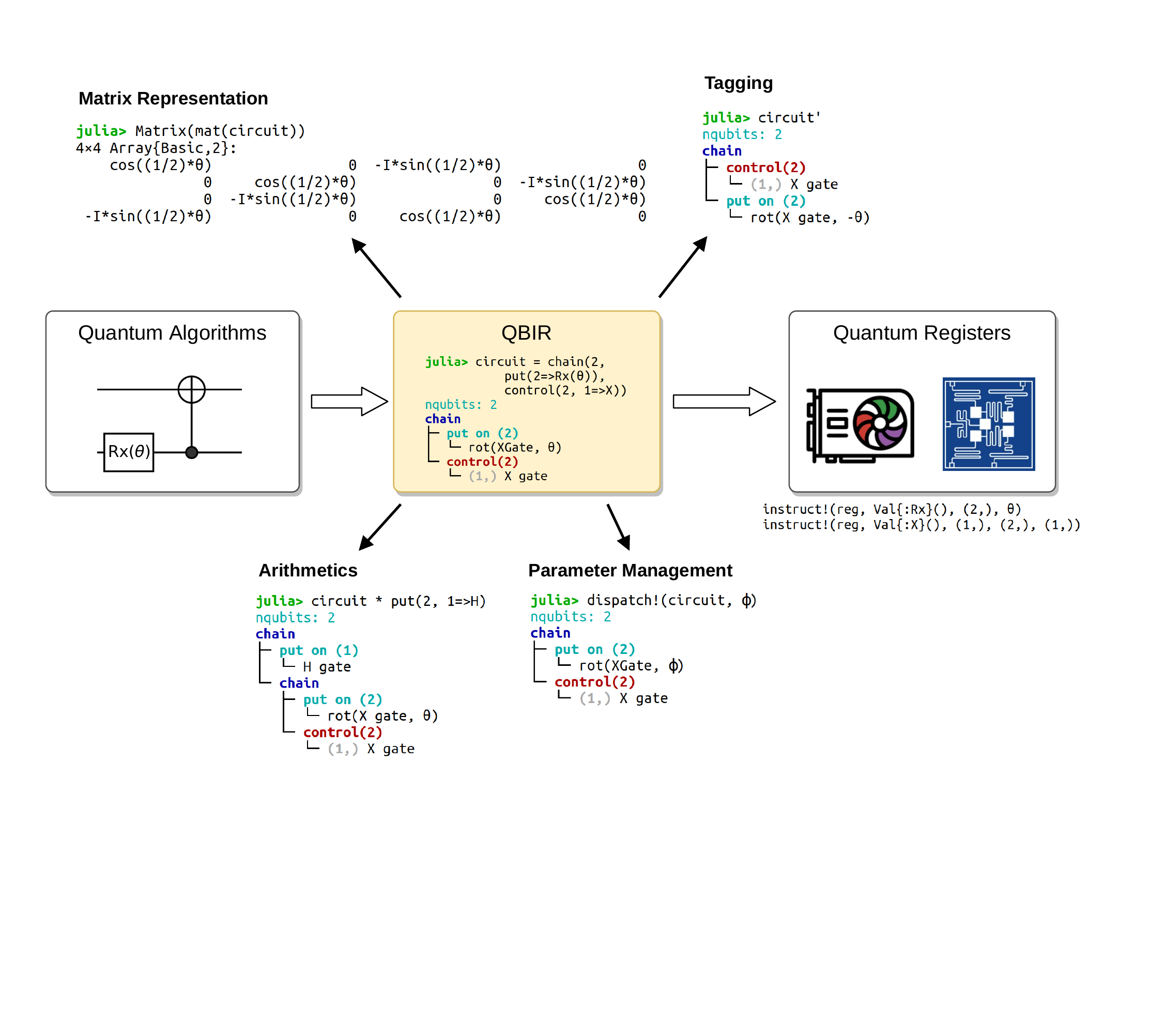}}
    \caption{
    Quantum block intermediate representation plays a central role in \texttt{Yao}. The images of GPU and quantum circuits are taken from JuliaGPU~\cite{besard2018effective} and IBM q-experience~\cite{Garcia2019}.  
    }\label{fig:arch}
\end{figure*}

On the other hand, deep learning and its extension \textit{differentiable programming} offer great inspiration and techniques for programming quantum computers.
Differentiable programming~\cite{diffprogramming} composes differentiable components to a learnable architecture and then learns the whole program by optimizing an objective function. The components are typically, but not limited to, neural networks. The word "differentiable" originates from the usual requirement of a gradient-based optimization scheme, which is crucial for scaling up to high dimensional parameter spaces. Differentiable programming removes laborious human efforts and sometimes produces even better programs than humans can produce themselves~\cite{karpathy}.

Differentiable programming is a sensible paradigm for variational quantum algorithms, where parameters of quantum circuits are modified within a particular parameter space, to optimize a loss function.
In this regard, programming quantum circuits in the differentiable paradigms address a much long term issue than the short-term considerations of compensating low-depth noisy quantum circuits with hybrid quantum-classical algorithms. Designing innovative and profitable quantum algorithms is, in general, nontrivial due to the lack of quantum intuitions. Fortunately, differentiable programming offers a new paradigm for devising novel quantum algorithms, much like what has already happened to the classical software landscape~\cite{karpathy}. 

The algorithmic advances in differentiable programming hugely benefit from rapid development in software frameworks~\cite{chen2015mxnet, abadi2016tensorflow, NEURIPS2019_9015, maclaurin2015autograd, Flux.jl-2018,innes2019zygote}, among which the automatic differentiation (AD) of the computational graph is the key technique behind the scene. A computational graph is a directed acyclic graph that models the computational process from input to output of a program. In order to evaluate gradients via the automatic differentiation, machine learning packages~\cite{chen2015mxnet, abadi2016tensorflow, NEURIPS2019_9015, maclaurin2015autograd,Flux.jl-2018,innes2019zygote} construct computational graphs in various ways.

It is instructive to view quantum circuits from the perspective of computational graphs with additional properties such as reversibility. In this regard, contextual analysis of the quantum computational graphs can be even more profitable than neural networks.
For example, uncomputing (also known as adjoint or dagger) a sub-program plays a central role in reversible computing~\cite{Bennett1973} since it returns qubit resources to the pool. While in differentiable programming of quantum circuits, exploiting the reversibility of the computational graph allows differentiating through the quantum circuit with constant memory independent to its depth.

Inspired by differentiable programming software, we design \texttt{Yao} to be around the domain-specific computational graph, the quantum block intermediate representation (QBIR). A block refers to a tensor representation of quantum operations, which can be quantum circuits and quantum operators of various granularities (quantum gates, Hamiltonian, or the whole program). As shown in \Fig{fig:arch}, QBIR offers a hardware-agnostic abstraction of quantum circuits. It is called an intermediate representation due to its stage in the quantum compilation, which bridges the high-level quantum algorithms and low-level device-specific instructions.  \texttt{Yao} provides rich functionalities to construct, inspect, manipulate, and differentiate quantum circuits in terms of QBIR. 

\begin{mdframed}[
    frametitle={What can \texttt{Yao} do ?},
    outerlinewidth=0.6pt,
    innertopmargin=6pt,
    innerbottommargin=6pt,
    roundcorner=4pt]    
   \begin{itemize}
    \item Optimize a variational circuit with $10,000$ layers using reverse-mode AD on a laptop, see \Lst{lst:ad}. 
    \item Construct sparse matrix representation of 20 site Heisenberg Hamiltonian in approximately 5 seconds, see \Lst{lst:benchmark-matrix-ad}.  
    \item Simulate Shor's 9 qubit error correction code symbolically, see \App{app:symbolic}.
    \item Send your circuit or Hamiltonian to a remote host in the form of \texttt{YaoScript}, see \App{app:yaoscript}.
    \item Compile an arbitrary two-qubit unitary to a target circuit structure via gradient optimization, see \App{app:gatelearning} and  \texttt{Yao}'s \href{https://github.com/QuantumBFS/QuAlgorithmZoo.jl/blob/v0.1.0/examples/PortZygote/gate\_learning.jl}{\texttt{QuAlgorithmZoo}}. 
    \item Solve ground state of a $6\times 6$ lattice spin model with a tensor network inspired quantum circuit on GPU~\cite{liu2019variational}.
    \end{itemize} 
\end{mdframed}\label{md:yaocando}

A distinct feature of \texttt{Yao} is its builtin automatic differentiation engine. Instead of building upon existing machine learning frameworks~\cite{chen2015mxnet, abadi2016tensorflow, NEURIPS2019_9015, maclaurin2015autograd, Flux.jl-2018,innes2019zygote}, we design \texttt{Yao}'s automatic differentiation engine to exploit reversibility in quantum computing, where the QBIR serves as a reversible computational graph. This implementation features speed and constant memory cost with respect to the circuit depth. 

\texttt{Yao} dispatches QBIR to low-level instructions of quantum registers of various types (CPU, GPU, and QPU in the future). Extensions of \texttt{Yao} can be done straightforwardly by defining new QBIR nodes or quantum register types. As a bonus of the generic design,  symbolic manipulation of quantum circuits in \texttt{Yao} follows almost for free. \texttt{Yao} achieves all these great flexibility and extensibility without sacrificing performance. \texttt{Yao} achieves one of the best performance for simulating quantum circuits of small to intermediate sizes Sec.\ref{sec:performance}, which are arguably the most relevant to quantum algorithms design for near-term devices.

\texttt{Yao} adds a unique solution to the landscape of open source quantum computing software includes \texttt{Quipper}~\cite{green2013quipper},
\texttt{ProjectQ}~\cite{steiger2016projectq},
\texttt{Q\#}~\cite{Svore2018},
\texttt{Cirq}~\cite{cirq},
\texttt{qulacs}~\cite{qulacs2019variational},
\texttt{PennyLane}~\cite{bergholm2018pennylane}, 
\texttt{qiskit}~\cite{Qiskit}, and \texttt{QuEST}~\cite{Jones2019}. References~\cite{fingerhuth2018open, LaRose2019overviewcomparison, Benedetti_2019} contain more complete surveys of quantum software. 
Most software represents quantum circuits as a sequence of
instructions. Thus, users need to define their abstraction for circuits with rich structures. \texttt{Yao} offers QBIR and related utilities to compose and manipulate complex quantum circuits. 
\texttt{Yao}'s QBIR is nothing but an abstract syntax tree, which is a commonly used data structure in modern programming languages thanks to its strong expressibility for control flows
and hierarchical structures. \texttt{Quipper}~\cite{green2013quipper} has adopted a similar strategy for the functional programming of quantum computing. \texttt{Yao} additionally introduces \texttt{Subroutine} to manage the scope of active and ancilla qubits. Besides these basic features, \texttt{Yao} puts a strong focus on differentiable programming of quantum circuits.
In this regards, \texttt{Yao}'s batched quantum register with GPU acceleration and built-in AD engine offers
significant speedup and convenience compared to \texttt{PennyLane}~\cite{bergholm2018pennylane} and \texttt{qulacs}~\cite{qulacs2019variational}.
    
\begin{figure}
    \centerline{\includegraphics[width=0.4\textwidth]{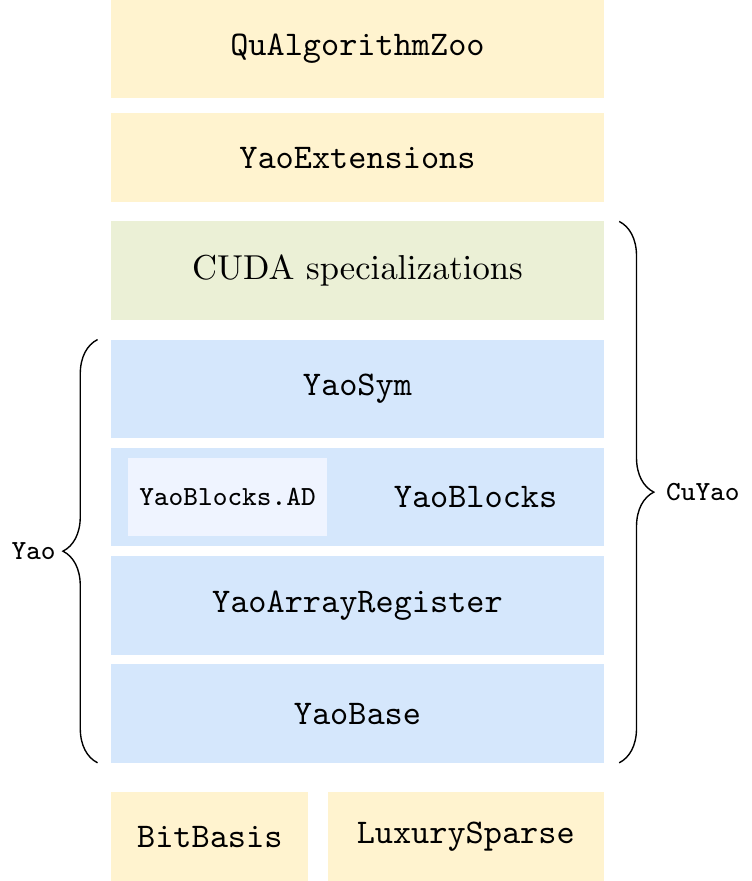}}
    \captionsetup{singlelinecheck=off}
    \caption[]{The packages in \texttt{Yao}'s ecosystem.  
    \begin{itemize}
        \item \textbf{BitBasis} provides bitwise operations,
        \item \textbf{LuxurySparse} is an extension of Julia's builtin \textbf{SparseArrays}. It defines customized sparse matrix types and implements efficient operations relevant to quantum computation. 
        \item \textbf{YaoBase} is an abstract package that defines basic abstract type for quantum registers and operations.
        \item \textbf{YaoArrayRegister} defines a register type as well as instruction sets,
        \item \textbf{YaoBlocks} defines utilities to construct and manipulate quantum circuits. It contains a builtin AD engine \textbf{YaoBlocks.AD} (Note: it is reexported in \textbf{Yao}, hence referred as \textbf{Yao.AD} in the main text).
        \item \textbf{YaoSym} provides symbolic computation supports.
        \item \textbf{Yao} is a meta package that re-export \textbf{YaoBase}, \textbf{YaoArrayRegister}, \textbf{YaoBlocks} and \textbf{YaoSym}.
        \item \textbf{CuYao} is a meta-package which contains \textbf{Yao} and provides specializations on CUDA devices.
        \item \textbf{YaoExtensions} provides utilities for constructing circuits and hamiltonians, faithful gradient estimator of quantum circuit, and some experimental features.
        \item \textbf{QuAlgorithmZoo} contains examples of quantum algorithms and applications.
    \end{itemize}}
    \label{fig:packages}
\end{figure}

An overview for the ecosystems of \texttt{Yao}, ranging from the low level customized bit operations and linear algebra to high-level quantum algorithms, is provided in Figure \ref{fig:packages}. In Sec.~\ref{sec:qbir} we introduce the quantum block intermediate representation; In Sec.~\ref{sec:ad} we explain the mechanism of the reversible computing and automatic differentiation in \texttt{Yao}. The quantum register which stores hardware specific information about the quantum states in \texttt{Yao} are explained in Sec.~\ref{sec:qregisters}. In Sec.~\ref{sec:performance}, we compare performance of \texttt{Yao} against other frameworks to illustrate the excellent efficiency of \texttt{Yao}. In Sec. \ref{sec:extending} we emphasize the flexibility and extensibility of Yao, perhaps its most important features for integrating with existing tools. The applications of \texttt{Yao} and future directions for developing \texttt{Yao} are discussed in Sec.~\ref{sec:application} and Sec.~\ref{sec:roadmap} respectively. Finally we summarize in Sec.~\ref{sec:sum}. The Appendices (\ref{app:external}-\ref{app:reading}) show various aspects and versatile applications of \texttt{Yao}. 

\begin{mdframed}[
    frametitle={Why Julia ?},
    outerlinewidth=0.6pt,
    innertopmargin=6pt,
    innerbottommargin=6pt,
    roundcorner=4pt]
Julia is fast! The language design avoids the typical compilation and execution uncertainties associated with dynamic languages~\cite{jeff2015juliacon}. Generic programming in Julia~\cite{bezanson2012julia}  
helps \texttt{Yao} reach optimized performance while still keeping the code base general and concise. 
Benchmarks in Sec.~\ref{sec:performance} show that \texttt{Yao} reaches one of the best performances with generic codes written purely in Julia. 

Julia codes can be highly extensible thanks to its type system and multiple dispatch mechanics. 
\texttt{Yao} builds its customized type system and dispatches to the quantum registers and circuits with a general interface. 
Moreover, Julia's meta-programming ability makes developing customized syntax and device-specific programs simple.
Julia's dynamic and generic approach for GPU programming~\cite{besard2018effective} powers \texttt{Yao}'s CUDA extension.

Julia integrates well with other programming languages.
It is generally straightforward to use external libraries written in other languages in \texttt{Yao}.
For example, the symbolic backend of \texttt{Yao} builds on \texttt{SymEngine} written in C++.
In \App{app:external}, we show an example of using 
the Python package \texttt{OpenFermion}~\cite{Mcclean2017} within \texttt{Yao}.
\end{mdframed}

\begin{figure}
    \centering
    \begin{tikzpicture}
    \tikzstyle{chain}=[draw, rectangle, rounded corners, fill=red!20, minimum width=1cm, minimum size=17pt]
    \tikzstyle{primitive}=[draw, rectangle, fill=green!20, minimum width=1cm, minimum size=17pt]
    \tikzstyle{ctrl}=[draw, circle, fill=black, minimum size=8, inner sep=0pt]
    \tikzstyle{composite}=[draw, rectangle, rounded corners, fill=red!20, dashed]
    \tikzstyle{line} = [thick,-,>=stealth]

    \node[composite, minimum width=4cm, minimum height=3.2cm] (chain-A) at (1.45, -0.9) {};
    \node[composite, minimum width=1.2cm, minimum height=2cm, fill=blue!20] (chain-B) at (1.2, -0.5) {};
    \node[composite, minimum width=1.2cm, minimum height=2.8cm, fill=blue!20] (chain-B) at (2.7, -0.9) {};

    \node[composite, minimum width=2.4cm, minimum height=2.2cm] (chain-A) at (4.88, -1.4) {};
    \node[composite, minimum width=1.2cm, minimum height=1.7cm, fill=blue!20] (chain-B) at (5.4, -1.4) {};

    \node[composite, minimum width=1cm, minimum height=1cm] (chain-A) at (6.8, -2) {};

    \node[composite, circle] (text-A) at (-0.3, 1) {};
    \node[right= 1.7 of text-A, anchor=east] (annotation-A) {\small hcphases};
    
    \node[composite, circle, fill=blue!20] (text-B) at (1.8, 1) {};
    \node[right= 1.5 of text-B, anchor=east] (annotation-B) {\small cphase};
    \node[draw, rectangle, fill=none, minimum size=17pt] (0, -4) {};

    \node[primitive] (H1) at (0, 0) {H};
    \node[primitive] (shift1) at (1.2, 0) {shift};
    \node[primitive] (shift2) at (2.7, 0) {shift};
    \node[ctrl] (ctrl1) at (1.2, -1) {};
    \node[ctrl] (ctrl2) at (2.7, -2) {};

    \node[primitive] (H2) at (4.2, -1) {H};
    \node[primitive] (shift3) at (5.4, -1) {shift};
    \node[ctrl] (ctrl3) at (5.4, -2) {};

    \node[primitive] (H3) at (6.8, -2) {H};

    \draw[line] (-1, 0) -- (H1) -- (shift1) -- (shift2) -- (7.5, 0);
    \draw[line] (-1, -1) -- (H2) -- (shift3) -- (7.5, -1);
    \draw[line] (-1, -2) -- (H3) -- (7.5, -2);

    \draw[line] (ctrl1) -- (shift1);
    \draw[line] (ctrl2) -- (shift2);
    \draw[line] (ctrl3) -- (shift3);
\end{tikzpicture}
    \caption{Quantum Fourier transformation circuit. The red and blue dashed blocks are built by the \textbf{hcphases} and \textbf{cphase} functions in the Listing~\ref{lst:qft}.}
    \label{fig:qft}
\end{figure}
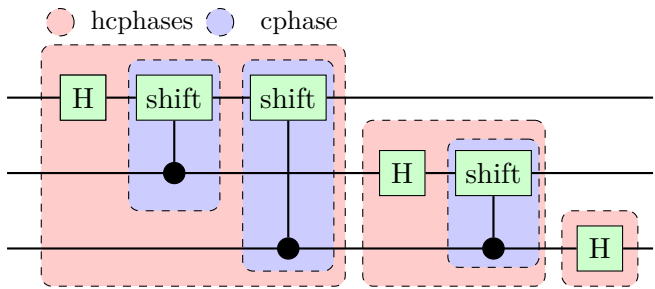

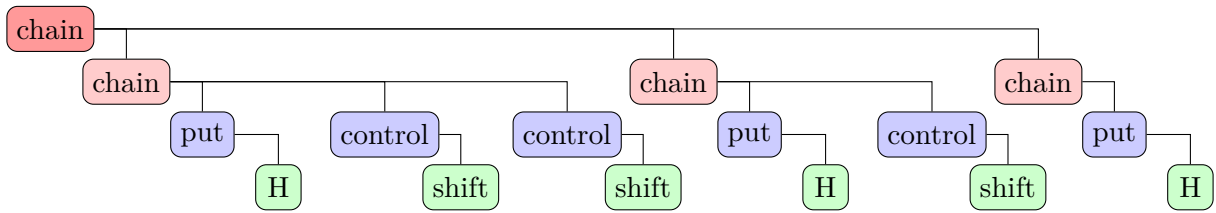
\begin{figure*}[t]
    \centering
    \begin{tikzpicture}[grow via three points={one child at (1.0, -0.7) and
    two children at (1.0, -0.7) and (2.2, -0.7)},
    edge from parent path={(\tikzparentnode.east) -| (\tikzchildnode.north)}]
    \tikzstyle{every node}=[draw, rectangle, rounded corners, fill=blue!20, minimum width=1cm, minimum size=17pt]
    \tikzstyle{chain}=[draw, rectangle, rounded corners, fill=red!20, minimum width=1cm, minimum size=17pt]
    \tikzstyle{primitive}=[draw, rectangle, rounded corners, fill=green!20, minimum width=1cm, minimum size=17pt]
    \tikzstyle{line} = [thick,-,>=stealth]
    \node[chain, fill=red!40] {chain}
        child {
            node[chain] {chain}
                child {
                    node {put}
                    child{ node[primitive] {H}}
                }
                child [missing] {}
                child {
                    node {control}
                    child{ node[primitive] {shift} }
                }
                child [missing] {}
                child {
                    node {control}
                    child{ node[primitive] {shift} }
                }
                child [missing] {}
        }
        child [missing] {}
        child [missing] {}
        child [missing] {}
        child [missing] {}
        child [missing] {}
        child {
            node[chain] {chain}
                child {
                    node {put}
                    child{ node[primitive] {H}}
                }
                child [missing] {}
                child {
                    node {control}
                    child{ node[primitive] {shift} }
                }
                child [missing] {}
        }
        child [missing] {}
        child [missing] {}
        child [missing] {}
        child {
            node[chain] {chain}
            child {
                    node {put}
                    child{ node[primitive] {H}}
                }
        };
\end{tikzpicture}
    \caption{Quantum Fourier transformation circuit as a QBIR. 
    The red nodes are roots of the composite \textbf{ChainBlock}. The blue nodes indicate the composite \textbf{ControlBlock} and \textbf{PutBlock}.  
    Green nodes are primitive blocks.}
    \label{fig:qft-tree}
\end{figure*}

\section{Quantum Block Intermediate Representation}\label{sec:qbir}
The QBIR is a domain-specific abstract syntax tree for quantum operators, including circuits and observables. In this section, we introduce QBIR and its central role in \texttt{Yao} via concrete examples. 

\subsection{Representing Quantum Circuits}\label{sec:representing-quantum-circuits}

Figure~\ref{fig:qft} shows the quantum Fourier transformation circuit~\cite{coppersmith1994approximate,ekert1996quantum, jozsa1998quantum} which contains the \texttt{hcphases} blocks (marked in red) of different sizes. Each block itself is also a composition of Hadamard gates and \texttt{cphase} blocks (marked in blue) on various locations. In \texttt{Yao}, it takes three lines of code to construct the QBIR of the QFT circuit.

\begin{minipage}{.44\textwidth}
\begin{lstlisting}[label=lst:qft, caption=quantum Fourier transform]
julia> using Yao

julia> cphase(i, j) = control(i, j=> shift(
    2π/(2^(i-j+1))));

julia> hcphases(n, i) = chain(n, i==j ?
   put(i=>H) : cphase(j, i) for j in i:n);

julia> qft(n) = chain(hcphases(n, i)
                for i in 1:n)

julia> qft(3)
nqubits: 3
chain
├─ chain
│  ├─ put on (1)
│  │  └─ H gate
│  ├─ control(2)
│  │  └─ (1,) shift(1.5707963267948966)
│  └─ control(3)
│     └─ (1,) shift(0.7853981633974483)
├─ chain
│  ├─ put on (2)
│  │  └─ H gate
│  └─ control(3)
│     └─ (2,) shift(1.5707963267948966)
└─ chain
   └─ put on (3)
      └─ H gate
\end{lstlisting}
\end{minipage}
The function \texttt{cphase} defines a control phase shift gate
with the \texttt{control} and \texttt{shift} functions. 
The function \texttt{hcphases} defines the recursive pattern in the QFT circuit, which puts a Hadamard gate in the first qubit of the subblock and then chains it with several control shift gates. The \texttt{chain} block is a composition of blocks with the same number of qubits. It is equivalent to matrix multiplication in reverse order mathematically.
Finally, one composes the QFT circuit of a given size by chaining the \texttt{hcphases} blocks. 
Overall, these codes construct a tree representation of the circuit shown in \Fig{fig:qft-tree}.
The subtrees are composite blocks (\texttt{ChainBlock}, \texttt{ControlBlock}, and \texttt{PutBlock}) with different
composition relations indicated in their roots. The leaves of the tree are primitive blocks. \App{app:block} shows the builtin block types of \texttt{Yao}, which are open to extension as shown in \App{app-qftblock}.  

In \texttt{Yao}, to execute a quantum circuit, one can simply feed a quantum state into the QBIR.

\begin{minipage}{.44\textwidth}
\begin{lstlisting}[label=lst:apply, caption=apply! and pipe] 
julia> rand_state(3) |> qft(3); 
     # same as apply!(rand_state(3), qft(3))
\end{lstlisting}
\end{minipage}

Here, we define a random state on 3 qubits and pass it through the QFT circuit.
The pipe operator \texttt{|>} is overloaded to call the \texttt{apply!} function
which applies the quantum circuit block to the register and modifies the register \textbf{inplace}.

The generic implementation of QBIR in \texttt{Yao} allows supporting both numeric and symbolic data types. 
For example, one can inspect the matrix representation of quantum gates defined in \texttt{Yao} with symbolic variables.

\begin{minipage}{.44\textwidth}
\begin{lstlisting}[label=lst:shift, caption=inspecting gates]
julia> using Yao, SymEngine

julia> @vars θ
(θ,)

julia> shift(θ) |> mat 
2×2 LinearAlgebra.Diagonal
{Basic,Array{Basic,1}}:
 1         ⋅
 ⋅  exp(im*θ)
 
julia> control(2,1,2=>shift(θ)) |> mat
4×4 LinearAlgebra.Diagonal{Basic,
                    Array{Basic,1}}:
 1  ⋅  ⋅         ⋅
 ⋅  1  ⋅         ⋅
 ⋅  ⋅  1         ⋅
 ⋅  ⋅  ⋅  exp(im*θ)
\end{lstlisting}
\end{minipage}
Here, the \texttt{@vars} macro declares the symbolic variable $\theta$.  The \texttt{mat} function constructs the matrix representation of a quantum block. \App{app:symbolic} shows another example of demonstrating Shor's 9 qubits code for quantum error correction with symbolic computation.  
 
\subsection{Manipulating Quantum Circuits}\label{sec:symbolic-manipulation}
In essence, QBIR represents the algebraic operations of a quantum circuit as types. Being an algebraic data type system, QBIR automatically allows pattern matching with Julia's multiple dispatch mechanics. Thus, one can manipulate quantum circuits in a straightforward manner using pattern matching on their QBIR.  

\begin{minipage}{.44\textwidth}
\begin{lstlisting}[label=lst:decompose, caption=gate decomposition]
julia> decompose(x::HGate) = 
    Rz(0.5π)*Rx(0.5π)*Rz(0.5π);
    
julia> decompose(x::AbstractBlock) =
    chsubblocks(x, decompose.(subblocks(x))); 
    
julia> qft(3) |> decompose
nqubits: 3
chain
├─ chain
│  ├─ put on (1)
│  │  └─ chain
│  │     ├─ rot(ZGate, 1.5707963267948966)
│  │     ├─ rot(XGate, 1.5707963267948966)
│  │     └─ rot(ZGate, 1.5707963267948966)
│  ├─ control(2)
│  │  └─ (1,) shift(1.5707963267948966)
│  └─ control(3)
│     └─ (1,) shift(0.7853981633974483)
├─ chain
│  ├─ put on (2)
│  │  └─ chain
│  │     ├─ rot(ZGate, 1.5707963267948966)
│  │     ├─ rot(XGate, 1.5707963267948966)
│  │     └─ rot(ZGate, 1.5707963267948966)
│  └─ control(3)
│     └─ (2,) shift(1.5707963267948966)
└─ chain
   └─ put on (3)
      └─ chain
         ├─ rot(ZGate, 1.5707963267948966)
         ├─ rot(XGate, 1.5707963267948966)
         └─ rot(ZGate, 1.5707963267948966)
\end{lstlisting}
\end{minipage}

For example, consider a practical situation where one needs to decompose the Hadamard gate into three rotation
gates~\cite{Karalekas_2020}. The codes in Listing~\ref{lst:decompose} define
compilation passes by dispatching the \texttt{decompose} function on different quantum block types.
For the generic \texttt{AbstractBlock},
we apply \texttt{decompose} recursively to all its sub-blocks and use the function \texttt{chsubblocks} defined in \texttt{Yao} 
to substitute the blocks.
The recursion terminates on primitive blocks where \texttt{subblocks} returns an empty set.
Due to the specialization of \texttt{decompose} method on Hadamard gates, a chain of three rotation gates are returned as a subblock instead.

\begin{minipage}{.44\textwidth}
\begin{lstlisting}[label=lst:iqft, caption=inverse QFT]
julia> iqft(n) = qft(n)';

julia> iqft(3)
nqubits: 3
chain
├─ chain
│  └─ put on (3)
│     └─ H gate
├─ chain
│  ├─ control(3)
│  │  └─ (2,) shift(-1.5707963267948966)
│  └─ put on (2)
│     └─ H gate
└─ chain
   ├─ control(3)
   │  └─ (1,) shift(-0.7853981633974483)
   ├─ control(2)
   │  └─ (1,) shift(-1.5707963267948966)
   └─ put on (1)
      └─ H gate
\end{lstlisting}
\end{minipage}

Besides replacing gates, one can also modify a block by applying tags to it. For example, the \texttt{Daggered} tag takes the hermitian conjugate of the block. We use the \texttt{\textquotesingle} operator to apply the \texttt{Daggered} tag. Similar to the implementation of \texttt{Transpose} on matrices in Julia, the dagger operator in \texttt{Yao} is "lazy" in the sense that one simply marks the block as \texttt{Daggered} unless there are specific daggered rules defined for the block. For example, the hermitian conjugate of a \texttt{ChainBlock} reverses the order of its child nodes and
propagate the  \texttt{Daggered} tag to each subblock. Finally, we have the following rules for primitive blocks, 
\begin{itemize}
    \item Hermitian gates are unchanged under dagger operation
    \item The hermitian conjugate of a rotational gate $R_{\sigma}(\theta) \rightarrow  R_{\sigma}(-\theta)$
    \item Time evolution block $e^{-iHt} \rightarrow e^{-iH(-t^*)}$
    \item Some special constant gates are hermitian conjugate to each other, e.g. \texttt{T} and \texttt{Tdag}.
\end{itemize}

With these rules, we can define the inverse QFT circuit directly in Listing~\ref{lst:iqft}.

\subsection{Matrix Representation} \label{sec:matrep}
Quantum blocks have a matrix representations of different types for optimized performance.
By default, the \texttt{apply!} method act quantum blocks to quantum registers using their matrix representations. 
The matrix representation is also useful for determining operator properties such as hermicity, unitarity, reflexivity, and commutativity. Lastly, one can also use \texttt{Yao}'s sparse matrix representation for quantum many-body computations such as exact diagonalization and (real and imaginary) time evolution. 

For example, one can construct the Heisenberg Hamiltonian and obtain its ground state using the Krylov space solver via the \texttt{KrylovKit.jl}~\cite{krylovkit} in Listing \ref{lst:heisenberg}. The arithmetic operations \texttt{*} and \texttt{sum} return \texttt{ChainBlock} and \texttt{Add} blocks respectively. It is worth noticing the differences between the QBIR arithmetic operations of the quantum circuits and those of Hamiltonians. Since the Hamiltonians are generators of quantum unitaries (i.e., $U = e^{-iHt}$), it is natural to perform additions for Hamiltonians (and other observables) and multiplications for unitaries. \texttt{YaoExtensions} provides some convenience functions for creating Hamiltonians on various lattices and variational quantum circuits.

\begin{minipage}{.44\textwidth}
\begin{lstlisting}[label=lst:heisenberg, caption=Heisenberg Hamiltonian]
julia> using KrylovKit: eigsolve

julia> bond(n, i) = sum([put(n, i=>σ) * put(
    n, i+1=>σ) for σ in (X, Y, Z)]);

julia> heisenberg(n) = sum([bond(n, i) 
            for i in 1:n-1]);
      
julia> h = heisenberg(16);
            
julia> w, v = eigsolve(mat(h)
              ,1, :SR, ishermitian=true)            
\end{lstlisting}
\end{minipage}

The \texttt{mat} function creates the sparse matrix representation of the Hamiltonian block. 
To achieve an optimized performance, we extend Julia's built-in sparse matrix types for various quantum gates. 
\App{app:sparse} lists these customized matrix types and promotion rules among them. 

Time evolution under a quantum Hamiltonian invokes the Krylov space method~\cite{DifferentialEquations.jl-2017}, which repeatedly applies the Hamiltonian block to the register. In this case, one can use the \texttt{cache} tag to create a \texttt{CachedBlock} for the Hamiltonian. Then, the \texttt{apply!} method makes use of the sparse matrix representation cached in the memory. 
Continuing from Listing~\ref{lst:heisenberg}, the following codes in Listing~\ref{lst:cache-te} show
that constructing and caching the matrix representation boosts the performance of time-evolution. 

\begin{minipage}{.44\textwidth}
\begin{lstlisting}[label=lst:cache-te, caption=Hamiltonian evolution is faster with cache]
julia> using BenchmarkTools

julia> te = time_evolve(h, 0.1);

julia> te_cache = time_evolve(cache(h), 0.1);

julia> @btime $(rand_state(16)) |> $te;
  1.042 s (10415 allocations: 1.32 GiB)

julia> @btime $(rand_state(16)) |> $te_cache;
  71.908 ms (10445 allocations: 61.48 MiB)
\end{lstlisting}
\end{minipage}

On the other hand, in many cases \texttt{Yao} can make use 
of efficient specifications of the \texttt{apply!} method for various blocks and apply them  
on the fly without generating the matrix representation.
The codes in Listing~\ref{lst:cache-qft} show that this approach can be faster for simulating quantum circuits.  

\begin{minipage}{.44\textwidth}
\begin{lstlisting}[label=lst:cache-qft, caption=Circuit simulation is faster without cache]
julia> r = rand_state(10);

julia> @btime r |> $(qft(10));
  550.466 μs (3269 allocations: 184.58 KiB)

julia> @btime r |> $(cache(qft(10)));
  1.688 ms (234 allocations: 30.02 KiB)
\end{lstlisting}
\end{minipage}

\section{Reversible Computing and Automatic Differentiation}\label{sec:ad}

Automatic differentiation efficiently computes the gradient of a program. It is the engine behind the success of deep learning~\cite{baydin2018automatic}. 
The technique is particularly relevant to differentiable programming of quantum circuits. 
In general, there are several modes of AD: the reverse mode caches intermediate state and then evaluate all gradients in a single backward run. The forward mode computes the gradients in a single pass together with the objective function, which does not require caching intermediate state but has to evaluate the gradients of all parameters one by one.

\texttt{Yao}'s builtin reverse mode AD engine (Sec.~\ref{sec:reverse-mode})
provides more efficient circuit differentiation for variational quantum algorithms compared to conventional reverse mode differentiation and forward mode differentiation (Sec. \ref{sec:forward-mode}).
By taking advantage of the reversible nature of quantum circuits, the memory complexity is reduced to constant compared to typical reverse mode AD~\cite{baydin2018automatic}. This property allows one to simulate very deep variational quantum circuits. Besides, \texttt{Yao} supports the forward mode AD (Sec.~\ref{sec:forward-mode}), which is a faithful quantum simulation of the experimental situation. In the classical simulation, the complexity of forward mode is unfavorable compared to reverse mode because one needs to run the circuit repeatedly for each component of the gradient.

\subsection{Reverse Mode: Builtin AD Engine with Reversible Computing} \label{sec:reverse-mode}
The submodule \texttt{Yao.AD} is a built-in AD engine. It back-propagates through quantum circuits using the computational graph information recorded in the QBIR.

In general, reverse mode AD needs to cache intermediate states in the forward pass for the backpropagation. Therefore, the memory consumption for backpropagating through a quantum simulator becomes unacceptable as the depth of the quantum circuit increases. Hence simply delegating AD to existing machine learning packages~\cite{chen2015mxnet, abadi2016tensorflow, NEURIPS2019_9015, maclaurin2015autograd,Flux.jl-2018,innes2019zygote} is not a satisfiable solution. 
\texttt{Yao}'s customized AD engine exploits the inherent reversibility of quantum circuits~\cite{griewank2008evaluating,gomez2017reversible}.  
By uncomputing the intermediate state in the backward pass, \texttt{Yao.AD} mostly performs in-place operations without allocations.    
\texttt{Yao.AD}'s superior performance is in line with the recent efforts of implementing efficient backpropagation through
reversible neural networks~\cite{gomez2017reversible,chen2018neural}.

In the forward pass we update the wave function $|\psi_k\rangle$ with inplace operations
\begin{align}
    \begin{split}
    \ldots\\
    |\psi_{k+1}\rangle = U_k |\psi_k\rangle ,\\
    \ldots
    \end{split}
\end{align}
where $U_k$ is a unitary gate parametrized by $\theta_k$. We define the adjoint of a variable as $\overline{x} = \frac{\partial \mathcal{L}}{\partial x^*}$ according to Wirtingers derivative~\cite{Hirose2003} for complex numbers, where $\mathcal{L}$ is a real-valued objective function that depends on the final state. Starting from $\overline{\mathcal{L}} = 1$ we can obtain the adjoint of the output state. 

To pull back the adjoints through the computational graph, we perform the backward calculation~\cite{Giles2008}
\begin{align}
    \begin{split}
        \ldots\\
        |\psi_k \rangle &=  U_k^\dagger | \psi_{k+1} \rangle \\
        \overline{|\psi_k \rangle} &=  U_k^\dagger \overline{| \psi_{k+1} \rangle}  \\
        \ldots
        \label{eq:apply-back}
    \end{split}
\end{align}
The two equations above are implemented \texttt{Yao.AD} with the \texttt{apply\_back!} method. Based on the obtained information, we can compute the  adjoint of the gate matrix using ~\cite{Giles2008}
\begin{align}
    \begin{split}
        \overline{U_k} = \overline{ | \psi_{k+1}  \rangle } \langle \psi_k|. 
        \label{eq:outer-product}
    \end{split}
\end{align}

This outer product is not  explicitly stored as a dense matrix. Instead, it is handled efficiently by customized low rank matrices described in \App{app:sparse}. 
Finally, we use \texttt{mat\_back!} method to compute the adjoint of gate parameters $\overline{\theta_k}$ 
from the adjoint of the unitary matrix $\overline{U_k}$. 

\begin{figure}
    \centerline{\includegraphics[width=\columnwidth,trim={2cm 1cm 2cm 0cm}, clip]{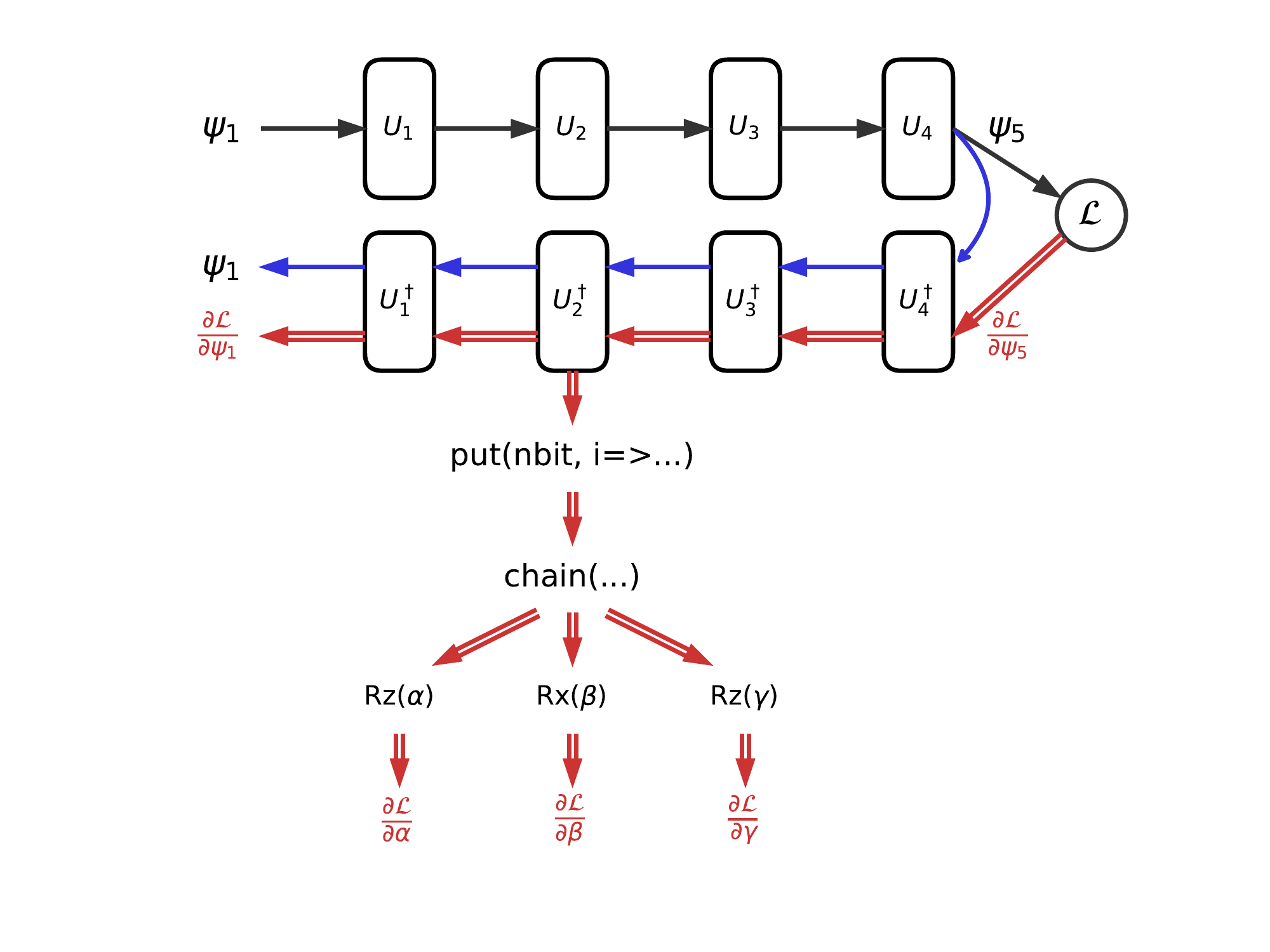}}
    \caption{Builtin automatic differentiation engine \textbf{Yao.AD}.
    Black arrows represent the forward pass. The blue arrow represents uncomputing. The red arrows indicate the backpropagation of the adjoints.}
    \label{fig:yaoad}
\end{figure}

Figure~\ref{fig:yaoad} demonstrates the procedure in a concrete example.
The black arrows show the forward pass without any allocation except for the output state and the objective function $\mathcal{L}$. 
In the backward pass, we uncompute the states (blue arrows) and backpropagate the adjoints (red arrows) at the same time. 
For the block defined as \texttt{put(nbit, i=>chain(Rz($\alpha$), Rx($\beta$), Rx($\gamma$)))}, 
we obtain the desired $\overline\alpha$, $\overline\beta$ and $\overline\gamma$
by pushing the adjoints back through the \texttt{mat} functions of \texttt{PutBlock} and \texttt{ChainBlock}.
The implementation of the AD engine is generic so that it works automatically with symbolic computation. 
We show an example of calculating the symbolic derivative of gate parameters in \App{app:symad}. One can also integrate 
\texttt{Yao.AD} with classical automatic differentiation engines such as \texttt{Zygote} to handle mixed classical and quantum computational graphs, see~\cite{betaVQE}. 

\begin{minipage}{.44\textwidth}
\begin{lstlisting}[label=lst:ad, caption=10000-layer VQE]
julia> using Yao, YaoExtensions

julia> n = 10; depth = 10000;

julia> circuit = dispatch!(
    variational_circuit(n, depth),
    :random);

julia> gatecount(circuit)
Dict{Type{#s54} where #s54 <:
    AbstractBlock,Int64} with 3 entries:
  RotationGate{1,Float64,ZGate} => 200000
  RotationGate{1,Float64,XGate} => 100010
  ControlBlock{10,XGate,1,1}    => 100000

julia> nparameters(circuit) 
300010

julia> h = heisenberg(n);

julia> for i = 1:100
    _, grad = expect'(h, zero_state(n)=>
                                circuit)
    dispatch!(-, circuit, 1e-3 * grad)
    println("Step $i, energy = $(expect(
            h, zero_state(10)=>circuit))")
       end
\end{lstlisting}
\end{minipage}

To demonstrate the efficiency of \texttt{Yao}'s AD engine, we use the codes in Listing~\ref{lst:ad} to simulate the variational quantum eigensolver (VQE)~\cite{Peruzzo2014} with depth $10,000$ (with $300,010$ variational parameters) on a laptop. The simulation would be extremely challenging without \texttt{Yao}, either due to overwhelming memory consumption in the reverse mode AD or unfavorable computation cost in the forward mode AD. 

Here, \texttt{variational\_circuit} is predefined in \texttt{YaoExtensions} to have a hardware efficient architecture~\cite{kandala2017hardware} shown in \Fig{fig:pcircuit-benchmark}. The \texttt{dispatch!} function with the second parameter specified to \texttt{:random} gives random initial parameters. 
The \texttt{expect} function
evaluates expectation values of the observables; the second argument can be a wave function or a pair of the input wave function and circuit ansatz like above.
\texttt{expect\textquotesingle} evaluates the gradient of this observable for the input wave function and circuit parameters. Here, we only make use of its second return value. For batched registers, the gradients of circuit parameters are accumulated rather than returning a batch of gradients.
\texttt{dispatch!(-, circuit, ...)} implements the gradient descent algorithm with energy as the loss function. The first argument is a binary operator that computes a new parameter based on the old parameter in \texttt{c} and the third argument, the gradients. Parameters in a circuit can be extracted by calling \texttt{parameters(circuit)}, which collects parameters into a vector by visiting the QBIR in depth-first order. The same parameter visiting order is used in \texttt{dispatch!}. In case one would like to share parameters in the variational circuit, one can simply use the same block instance in the QBIR. In the training process, gradients can be updated in the same field. After the training, the circuit is fully optimized and returns the ground state of the model Hamiltonian with zero state as input.

\subsection{Forward Mode: Faithful Quantum Gradients} \label{sec:forward-mode}
Compared to the reverse mode, forward mode AD is more closely related to how one measures the gradient in the actual experiment.

The implementation of the forward mode AD is particularly simple for the ``rotation gates'' $R_\Sigma(\theta) \equiv e^{-i\Sigma\theta/2}$
with the generator $\Sigma$ being both hermitian and reflexive ($\Sigma^2 = 1$). For example, $\Sigma$ can be the Pauli gates  X,
Y and Z, or multi-qubit gates such as CNOT, CZ, and SWAP. Every two-qubit gate can be decomposed into Pauli rotations and CNOTs (or CZs) via gate transformation~\cite{Crooks2019}. Under these conditions, the gradient to a circuit parameter is~\cite{Li2017,mitarai2018quantum, Schuld2019, Nakanishi2019}
\begin{equation}
    \frac{\partial \langle O \rangle_\theta}{\partial \theta} = \frac{1}{2}\left(\langle O \rangle_{\theta+\frac{\pi}{2}} - \langle O \rangle_{\theta-\frac{\pi}{2}}\right)
    \label{eq:shiftrule}
\end{equation}
where $\langle O \rangle_\theta$ denotes the expectation of the observable $O$ with the given parameter $\theta$. Therefore, one just needs to run the simulator twice to estimate the gradient. \texttt{YaoExtensions} implements \Eq{eq:shiftrule} with Julia's broadcasting semantics and obtains the full gradients with respect to all parameters. Similar features can be found in \texttt{PennyLane}~\cite{bergholm2018pennylane} and \texttt{qulacs}~\cite{qulacs2019variational}. We refer this approach as the \textit{faithful gradient}, since it mirrors the experimental procedure on a real quantum device. In this way, one can estimate the gradients in the VQE example Listing~\ref{lst:ad} using \Eq{eq:shiftrule}

\begin{minipage}{.44\textwidth}
\begin{lstlisting}
# this will be slow 
julia> grad = faithful_grad(h, zero_state(n)
                =>circuit; nshots=100);
\end{lstlisting}
\end{minipage}
where one faithfully simulates \texttt{nshots} projective measurements. 
In the default setting \texttt{nshots=nothing}, the function evaluates the exact expectation on the quantum state. Note that simulating projective measurement, in general, involves rotating to eigenbasis of the observed operator. \texttt{Yao} implements an efficient way to break the measurement into the expectation of local terms by diagonalizing the observed operator symbolically as bellow.

\begin{minipage}{.44\textwidth}
\begin{lstlisting}[label=lst:eigen, caption=The eigendecomposition of a QBIR.]
julia> O = chain(5, put(5,2=>X), put(5,3=>Y))
nqubits: 5
chain
├─ put on (2)
│  └─ X
└─ put on (3)
   └─ Y

julia> E, U = YaoBlocks.eigenbasis(O)
(nqubits: 5
chain
├─ put on (2)
│  └─ Z
└─ put on (3)
   └─ Z
, nqubits: 5
chain
├─ put on (2)
│  └─ H
└─ put on (3)
   └─ chain
      ├─ H
      └─ S
)
\end{lstlisting}
\end{minipage}

The return value of \texttt{eigenbasis} contains two QBIRs \texttt{E} and \texttt{U} such that \texttt{O = U*E*U\textquotesingle}.
\texttt{E} is a diagonal operator that represents the observable in the measurement basis. \texttt{U} is a circuit that rotates computational basis to the measurement basis.

The above gradient estimator \Eq{eq:shiftrule} can also be generalized to statistic functional loss, which is useful for generative modeling with an implicit probability distribution given by the quantum circuits~\cite{liu2018differentiable}. The symmetric statistic functional of order two reads
\begin{equation}
    \mathcal{F}_\theta = \expect{K(x, y)}{x\sim p_\theta, y\sim p_\theta},
\end{equation}
where $K$ is a symmetric function, $p_\theta$ is the output probability distribution of a parametrized quantum circuit measured on the computational basis. If the circuit is parametrized by rotation gates, the gradient of the  statistic functional is
\begin{eqnarray}
    \frac{\partial \mathcal{F}_\theta}{\partial \theta} =&\expect{K(x,y)}{x\sim p_{\theta + \frac{\pi}{2}}, y\sim p_\theta}\nonumber\\&-\expect{K(x,y)}{x\sim p_{\theta-\frac{\pi}{2}},y\sim p_\theta}, 
\end{eqnarray}
which is also related to the measure valued gradient estimator for stochastic optimization~\cite{Mohamed2019}.
Within this formalism, \texttt{Yao} provides the following interfaces to evaluate gradients with respect
to the maximum mean discrepancy loss~\cite{Li2017e, Gretton2012}, which measures the probabilistic distance between two sets of samples.

\begin{minipage}{.44\textwidth}
\begin{lstlisting}[label=lst:mmd, caption=gradient of maximum mean discrepancy]
julia> target_p = normalize!(rand(1<<5));

julia> kf = brbf_kernel(2.0);

julia> circuit = variational_circuit(5);

julia> mmd = MMD(kf, target_p);

julia> g_reg, g_params = expect'(
    mmd, zero_state(5)=>circuit);

julia> g_params = faithful_grad(
    mmd, zero_state(5)=>circuit);
\end{lstlisting}
\end{minipage}

\section{Quantum Registers}\label{sec:qregisters} 
The quantum register stores hardware-specific information about the quantum states.  
In classical simulation on a CPU, the quantum register is an array containing the quantum wave function. 
For GPU simulations, the quantum register stores the pointer to a GPU array. In an actual experiment, the register should be the quantum device that hosts the quantum state. \texttt{Yao} handles all of these cases with a unified \texttt{apply!} interface, which dispatches the instructions depending on different types of QBIR nodes and registers. 

\subsection{Instructions on Quantum Registers} \label{sec:instruct}

Quantum registers store quantum states in contiguous memory, which can either 
be the CPU memory or other hardware memory, such as a CUDA device.

\begin{minipage}{.44\textwidth}
\begin{lstlisting}[label=lst:cuyao, caption=CUDA register]
julia> using CuYao

# construct the |1010> state
julia> r = ArrayReg(bit"1010");

# transfer data to CUDA
julia> r = cu(r);
\end{lstlisting}
\end{minipage}

Each register type has its own device-specific instruction set. They are declared in \texttt{Yao} via the "instruction set" interface, which includes
\begin{itemize}
    \item \textbf{gate instruction}: \texttt{instruct!}
    \item \textbf{measure instruction}: \texttt{measure} and \texttt{measure!}
    \item \textbf{qubit management instructions}: \texttt{focus!} and \texttt{relax!}
\end{itemize}
The instruction interface provides a clean way to extend support to various backends without the user having to worry about changes to frontend interfaces.

\begin{minipage}{.44\textwidth}
\begin{lstlisting}[label=lst:instruct, caption=instruct! and measure]
julia> r = zero_state(4); 

julia> instruct!(r, Val(:X), (2, )) 
ArrayReg{1, Complex{Float64}, Array...}
    active qubits: 4/4
    
julia> samples = measure(r; nshots=3)
3-element Array{BitBasis.BitStr{4,Int64},1}:
 0010 ₍₂₎
 0010 ₍₂₎
 0010 ₍₂₎

julia> [samples[1]...] 
4-element Array{Int64,1}:
 0
 1
 0
 0
\end{lstlisting}
\end{minipage}

For example, the rotation gate shown in \Fig{fig:arch} is interpreted as
\texttt{instruct!(reg, Val(:Rx), (2,), $\theta$)}. The second parameter specifies
the gate, which is a \texttt{Val} type with a gate symbol as a type parameter. The third parameter is
the qubit to apply, and the fourth parameter is the rotation angle. The CNOT gate is interpreted as \texttt{instruct!(reg, Val(:X), (1,), (2,), (1,))}, where the last three
tuples are gate locations, control qubits, and configuration of the control qubits (0 for inverse control, 1 for control). Respectively.
The \texttt{measure} function simulates measurement from the quantum register and provides bit strings,
while \texttt{measure!} returns the bit string and also collapse the state. 

In the last line of the above example, we convert a bit string $\texttt{0010}_{\texttt{(2)}}$ to a vector \texttt{[0, 1, 0, 0]}. Note that the order is reversed since the readout of a bit string is in the little-endian format.

\subsection{Active qubits and environment qubits}\label{sec:scoping}
\begin{figure}[h]
    \centerline{\includegraphics[trim={0 1.2cm 0 0}, clip, width=9cm]{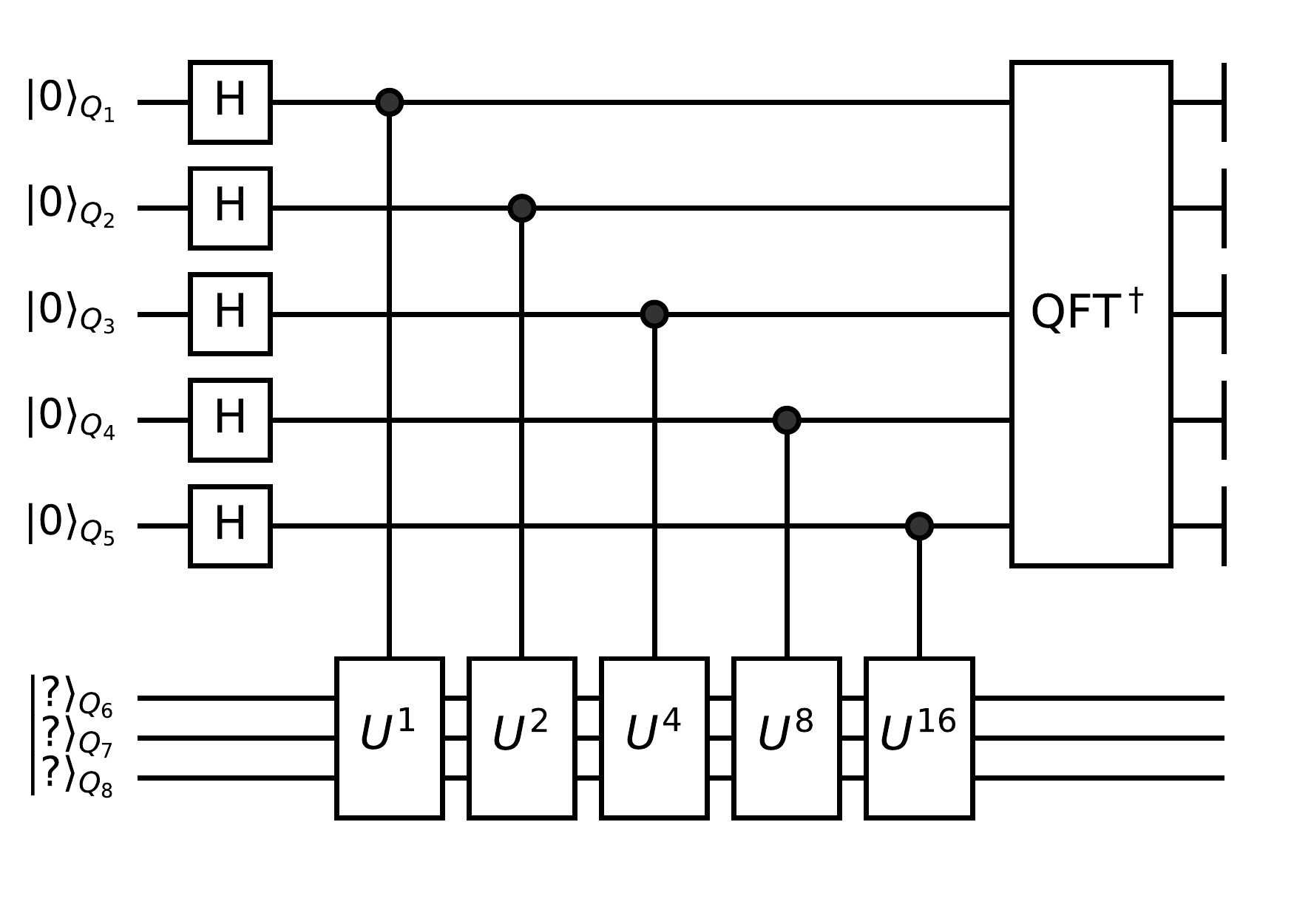}}
    \caption{5-qubit quantum Phase estimation circuit. This circuit contains three components.
    First, apply Hadamard gates to $n$ ancilla qubits. Then apply controlled unitary to $n+m$ qubits and finally apply inverse QFT to $n$ ancilla qubits.}\label{fig:phase-estimation}
\end{figure}

In certain quantum algorithms, one only applies the circuit block to a subset of qubits. For example, see the quantum phase estimation ~\cite{nielsen2010quantum} shown in \Fig{fig:phase-estimation}.\\

The QFT circuit block defined in Listing~\ref{lst:qft} can not be used directly in this case since the block size does not match the number of qubits. We introduce the concept of active and environment qubits to address this issue. Only the active qubits are visible to circuit blocks under operation. We manage the qubit resources with the \texttt{focus!} and its reverse \texttt{relax!} instructions.
For example, if we want to apply the QFT algorithm on qubits 3,6,1 and 2,
the \texttt{focus!} activates the four qubits 3,6,1,2 in the given order and deactivates the rest

\begin{minipage}{.44\textwidth}
\begin{lstlisting}[label=lst:focusrelax, caption=focus! and relax!]
julia> reg = rand_state(10)

julia> focus!(reg, (3,6,1,2))

julia> reg |> qft(4)

julia> relax!(reg, (3,6,1,2); to_nactive=10)
\end{lstlisting}
\end{minipage}

Since it is a recurring pattern to first \texttt{focus!}, then \texttt{relax!} on the same qubits in many quantum algorithms, we introduce a \texttt{Subroutine} node to manage the scope automatically. 
Hence, the phase estimation circuit in \Fig{fig:phase-estimation} can be defined with the following codes.

\begin{minipage}{.44\textwidth}
\begin{lstlisting}[label=lst:pe, caption=quantum phase estimation]
PE(n, m, U) = chain(
    n+m, # total number of qubits
    repeat(H, 1:n), # apply H from 1:n
    chain(control(
        k,
        n+1:n+m=>matblock(U^(2^(k-1))))
        for k in 1:n    
    ),
    
    # apply inverse QFT on a local scope
    subroutine(qft(n)', 1:n)
)
\end{lstlisting}
\end{minipage}
The \texttt{matblock} method in the codes constructs a quantum circuit from a given unitary matrix.

\subsection{Batched Quantum Registers}\label{sec:batched}

The batched register is a collection of quantum wave functions. It can be samples of classical data for quantum machine learning tasks~\cite{huggins2018towards} or an ensemble of pure quantum states for thermal state simulation~\cite{betaVQE}. For both applications, having the batch dimension not only provides convenience but may also significantly speed up the simulations.

We adopt the Single Program Multiple Data (SPMD)~\cite{darema1988single} design in \texttt{Yao} similar to modern machine learning frameworks so that it can make use of modern multi-processors such as multi-threading or GPU support
(and potentially multi-processor QPUs).
Applying a quantum circuit to a batched register means to apply the
same quantum circuit to a batch of wave functions in parallel,
which is extremely friendly to modern multi-processors.

The memory layout of the quantum register is a matrix of the size $2^a \times 2^r B$, 
where $a$ is the number of system qubits, $r$ is the number of remaining qubits (or environment qubits), $B$ is the batch size. For gates acting on the active qubits, the remaining qubits and batch dimension can be treated on an equal footing.
We put the batch dimension as the last dimension because Julia array is column majored. As the last dimension, it favors broadcasting on the batch dimensions.

One can construct a batched register in \texttt{Yao} and perform operations on it. These operations are automatically broadcasted over the batch dimension. 

\begin{minipage}{.44\textwidth}
\begin{lstlisting}[label=lst:batched-reg, caption=a batch of quantum registers]
julia> reg = rand_state(4; nbatch=5);

julia> reg |> qft(4) |> measure!
5-element Array{BitBasis.BitStr{4,Int64},1}:
 1011 ₍₂₎
 1011 ₍₂₎
 0000 ₍₂₎
 1101 ₍₂₎
 0111 ₍₂₎
\end{lstlisting}
\end{minipage}
Note that we have used the \texttt{measure!} function to collapse all batches. 

The measurement results are represented in \texttt{BitStr} type which is a subtype of \texttt{Integer} and has a static length. Here, it pretty-prints the measurement results and provides a convenient readout of measuring results.

\section{Performance}\label{sec:performance}

As introduced above, \texttt{Yao} features a generic and extensible implementation without sacrificing performance. Our performance optimization strategy heavily relies on Julia's multiple dispatch. As a bottom line, \texttt{Yao} implements a general multi-control multi-qubit arbitrary-location gate instruction as the fallback. We then fine-tune various specifications for better performance.  Therefore, in many applications, the construction and operation of QBIR do not even invoke matrix allocation. While in cases where the gate matrix is small (number of qubits smaller than 4), \texttt{Yao} automatically employs the corresponding static sized types~\cite{staticarrays} for better performance.
The sparse matrices \texttt{IMatrix}, \texttt{Diagonal}, \texttt{PermMatrix} and \texttt{SparseMatrixCSC} introduced in \App{app:sparse} also have their static version defined in \texttt{LuxurySparse.jl}~\cite{luxurysparse}.
Besides, we also utilize unique structures of frequently used gates and dispatch to specialized implementations. For example, 
Pauli X gate can be executed by swapping the elements in
the register directly.

\begin{figure*}[t]
\centerline{\includegraphics[width=0.95\textwidth]{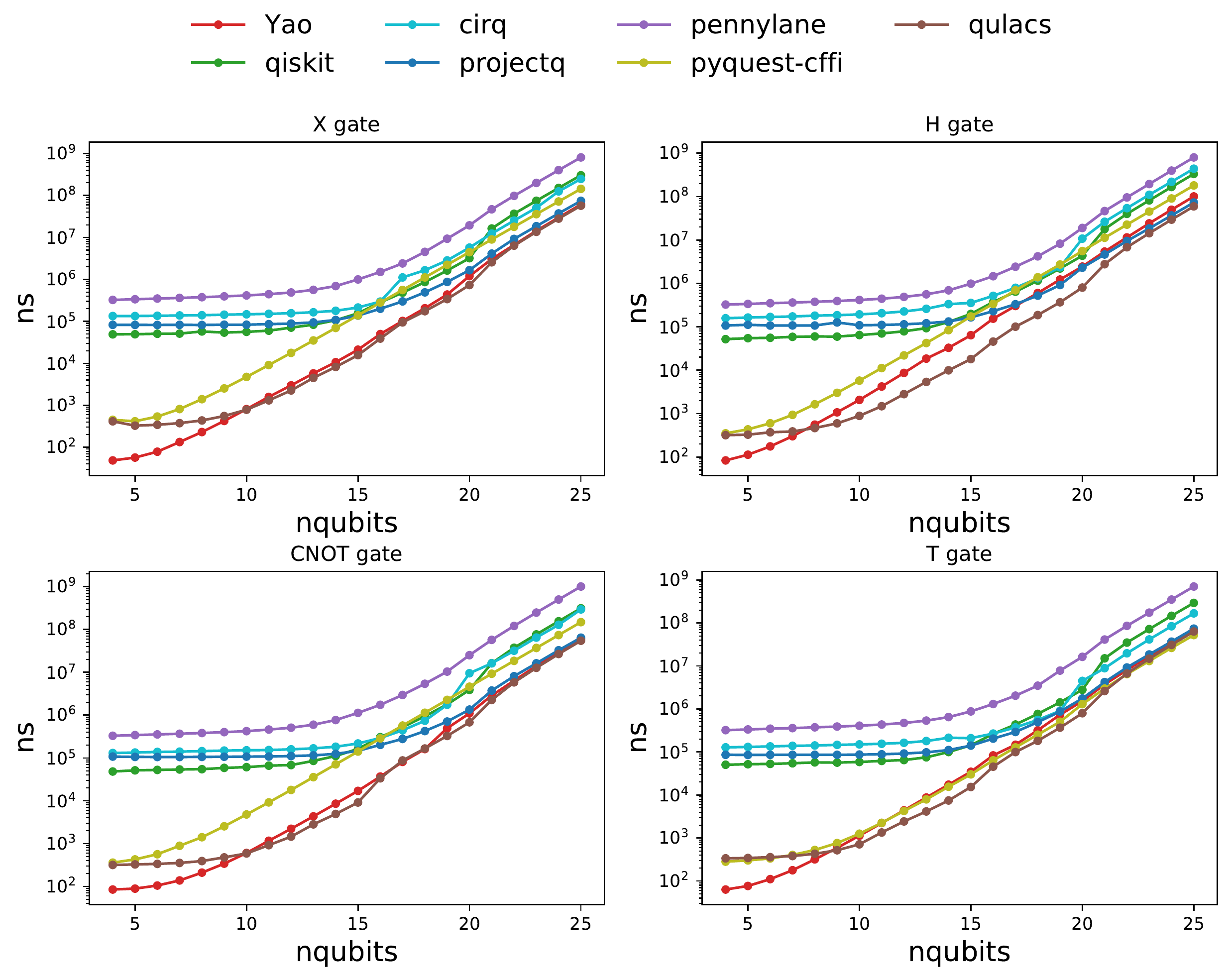}}
    \caption{Benchmarks of (a) Pauli-X gate; (b) Hadamard gate; (c) CNOT gate; (d) Toffolli gate.}
  \label{fig:benchmark}
\end{figure*}

\begin{figure*}[t]
    \centerline{\includegraphics[width=1\textwidth]{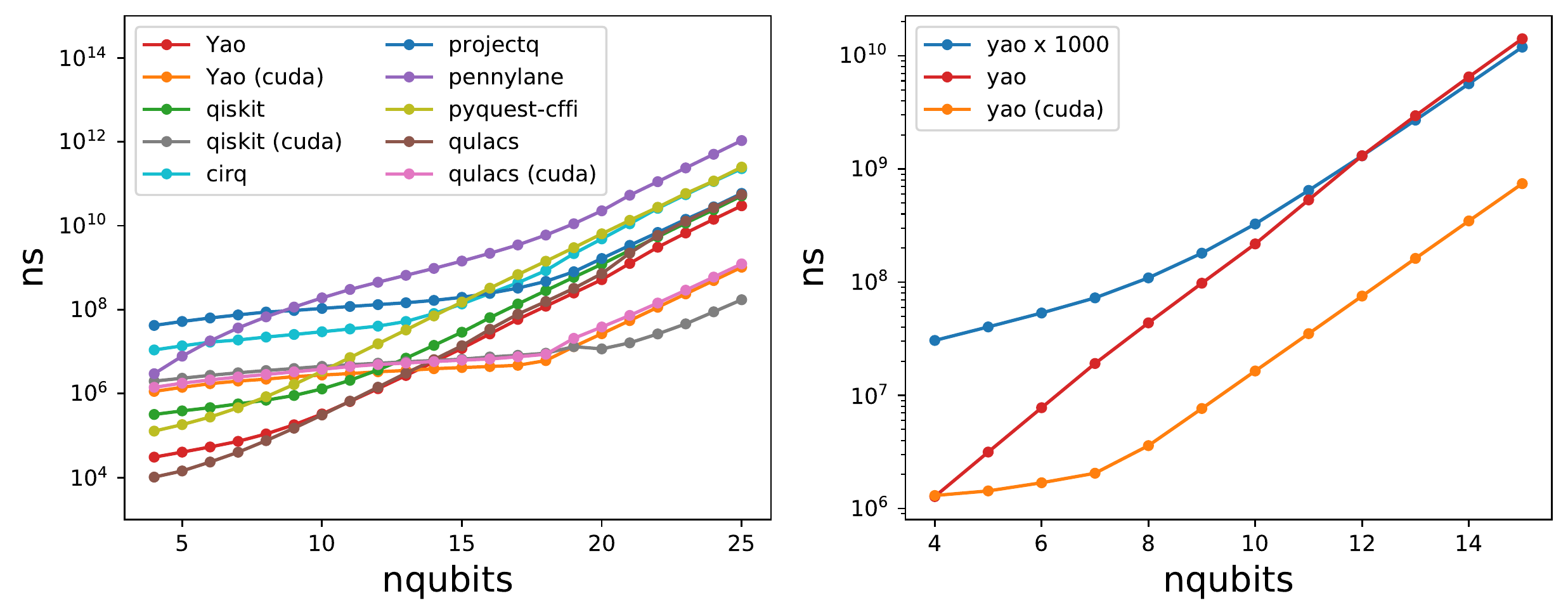}}
    \caption{
        (a) A parameterized quantum circuit with single qubit rotation and CNOT gates;
        (b) Benchmarks of the parameterized circuit;
        (c) Benchmarks of the parametrized circuit, the batched version. Line ``yao" represents the batched registers, ``yao (cuda)" represents the batched register on GPU, ``yao $\times$ 1000" is running on a non-batched register repeatedly for $1000$ times.
    }
  \label{fig:pcircuit-benchmark}
\end{figure*}

We benchmark \texttt{Yao}'s performance with other quantum computing software. Note that the exact classical simulation of the generic quantum circuit is doomed to be exponential~\cite{haner20170,markov2008simulating, pednault2017breaking, zhang2019alibaba}. \texttt{Yao}'s design puts a strong emphasis on the performance of small to intermediate-sized quantum circuits since the high-performance simulation of such circuits is crucial for the design of near-term algorithms that run repeatedly or in parallel.

\subsection{Experimental Setup}

\begin{table}[h!]\centering
\begin{minipage}{\columnwidth}
\ra{1.3}
    \scalebox{1.0}{
        \begin{tabularx}{\textwidth}{X X c}\toprule
            \textbf{Package} & \textbf{Language} & \textbf{Version}\\
            \hline
            \texttt{Cirq~\cite{cirq}} & Python & 0.8.0\\
            \texttt{qiskit~\cite{Qiskit}} & C++/Python & 0.19.2\\
            \texttt{qulacs~\cite{qulacs2019variational}} & C++/Python & 0.1.9\\
            \texttt{PennyLane~\cite{bergholm2018pennylane}} & Python & 0.7.0\\
            \texttt{QuEST~\cite{Jones2019}} & C/Python & 3.0.0\\
            \texttt{ProjectQ~\cite{steiger2016projectq}} & C++/Python & 0.4.2\\
            \texttt{Yao} & Julia & 0.6.2\\
            \texttt{CuYao} & Julia & 0.2.2\\
            \bottomrule
        \end{tabularx}
    }
    \caption{Packages in the benchmark.}\label{tbl-mattype}
\end{minipage}
\end{table}
Although \texttt{QuEST} is a package originally written in C,
we benchmark it in Python via \texttt{pyquest-cffi}~\cite{pyquest} for convenience.
\texttt{Pennylane} is benchmarked with its default backend~\cite{pennylane2019repo}. Since the package was designed primarily for being run on the quantum hardware, its benchmarks contain a certain overhead that was not present in other frameworks~\cite{pennylane2019issue}. \texttt{qiskit} is benchmarked with \texttt{qiskit-aer} 0.5.1~\cite{qiskit2019aer} and \texttt{qiskit-terra} 0.14.1~\cite{qiskit2019terra} using the statevector method of the qasm simulator.

\begin{table}[h!]\centering
\begin{minipage}{\columnwidth}
\ra{1.3}
    \scalebox{1.1}{
        \begin{tabularx}{0.9\textwidth}{X c}\toprule
            \textbf{Software} & \textbf{Version}\\
            \hline
            Python & 3.8.3\\
            Numpy & 1.18.1\\
            MKL & 2019.3\\
            Julia & 1.5.2\\
            \bottomrule
        \end{tabularx}
    }
    \caption{The environment setup of the machine for benchmark.}\label{tbl-env}
\end{minipage}
\end{table}

Our test machine contains an Intel(R) Xeon(R) Gold 6230 CPU with a Tesla V100 GPU accelerator.
SIMD is enabled with \textbf{AVX2} instruction set. The benchmark time is measured via \texttt{pytest-benchmark}~\cite{pytest2019benchmark} and
\texttt{BenchmarkTools}~\cite{BenchmarkTools.jl-2016} with minimum running time. We ignore the compilation time
in Julia since one can always get rid of such time by compiling the program ahead of time. The benchmark scripts and complete reports are maintained online at the repository~\cite{quantum2019benchmark}. For more detailed and latest benchmark configuration one should always refer to this repository.

\subsection{Single Gate Performance}\label{sec:single-gate-performance}
We benchmark several frequently used quantum gates, including the Pauli-X Gate, the Hadamard gate (H), the controlled-NOT gate (CNOT), and the Toffoli Gate. These benchmarks measure the performance of executing one single gate instruction.

Figure~\ref{fig:benchmark} shows the running times of various gates applied on the second qubit of the register from size 4 to 25 qubits in each package in the unit of nano seconds. One can see that \texttt{Yao}, \texttt{ProjectQ}, and \texttt{qulacs} reach similar performance when the number of qubits $n>20$. 
They are at least several times faster than other packages.
Having similar performance in these three packages suggests that they all reached the top performance for this type of full amplitude classical simulation on CPU.

\subsection{Parametrized Quantum Circuit Performance}\label{sec:circuit-performance}
Next, we benchmark the parameterized circuit of depth $d=10$ shown in \Fig{fig:pcircuit-benchmark}(a). This type of hardware-efficient circuits was employed in the VQE experiment~\cite{kandala2017hardware}. These benchmarks further test the performance of circuit abstraction in practical applications. 

The results in \Fig{fig:pcircuit-benchmark}(b) shows that \texttt{Yao} reaches the best performance for more than 10 qubits on CPU. 
\texttt{qulacs}'s well tuned C++ simulator is faster than \texttt{Yao} for fewer qubits. 
On a CUDA device, \texttt{Yao} and \texttt{qulacs} show similar performance. \texttt{qiskit} cuda backend shows better performance for more that 20 qubits. These benchmarks 
also, show that CUDA parallelization starts to be beneficial for a qubit number larger than $16$.
Overall, \texttt{Yao} is one of the fastest quantum circuit simulators for this type of application. 

Lastly, we benchmark the performance of batched quantum register introduced in Sec \ref{sec:batched}
in \Fig{fig:pcircuit-benchmark}(c) with a batch size $1000$. 
We only measure \texttt{Yao}'s performance due to the lack of native support of SPMD
in other quantum simulation frameworks. \texttt{Yao}'s CUDA backend (labeled as \texttt{yao (cuda)}) offers large speed up (>10x) compared to the CPU backend (labeled as \texttt{yao}). For reference, we also plot the timing of a bare loop over the batch dimension on a CPU (labeled as \texttt{yao $\times$ 1000}). One can see that batching offers substantial speedup for small circuits. 

The overhead of simulating small to intermediate-sized circuits is particularly relevant for designing variational quantum algorithms where the same circuit may be executed million times during training. \texttt{Yao} shows the least overhead in these benchmarks.
\texttt{qulacs} also did an excellent job of suppressing these overheads. 

\subsection{Matrix Representation and Automatic Differentiation Performance}\label{sec:mat-performance}

As discussed in Sec.~\ref{sec:matrep} and Sec.~\ref{sec:reverse-mode}, \texttt{Yao} features highly optimized matrix representation and reverse mode automatic differentiation for the QBIR. We did not attempt a systematic benchmark due to the lack of similar features in other quantum software frameworks. 

Here, we simply show the timings of constructing the sparse matrix representation of 20 site Heisenberg Hamiltonian and differentiating its energy expectation through a variational quantum circuit of depth 20 (200 parameters) on a laptop. The forward mode AD discussed in Sec.~\ref{sec:forward-mode} is slower by order of a hundred in such simulations. 

\begin{minipage}{0.44\textwidth}
\begin{lstlisting}[label=lst:benchmark-matrix-ad, caption=benchmark mat and AD performance]
julia> using BenchmarkTools, Yao, 
                    YaoExtensions

julia> @btime mat($(heisenberg(20)));
  6.330 s (10806 allocations: 10.34 GiB)

julia> @btime expect'($(heisenberg(20)), 
                      $(zero_state(20))=> 
              $(variational_circuit(20)));
  5.054 s (58273 allocations: 4.97 GiB)
\end{lstlisting}
\end{minipage}

\section{Extensibility}\label{sec:extending}
In the previous section we have demonstrated the excellent efficiency of \texttt{Yao} in comparison to other frameworks. We nevertheless emphasize that the most important feature of \texttt{Yao} is its flexibility and extensibility.
\subsection{Extending QBIR nodes}\label{sec:extend-ir-nodes}

It is easy to extend \texttt{Yao} with new gates and quantum block nodes. One can define constant gates by giving its matrix representation. For example, the \texttt{FSim} gate that appears in Google supremacy experiments~\cite{Google2019} is composed of an \texttt{ISWAP} gate and a \texttt{cphase} gate with a fixed angle. 
\begin{minipage}{0.44\textwidth}
\begin{lstlisting}[label=lst:fsim, caption=FSim gate]
julia> using Yao, LuxurySparse

julia> @const_gate ISWAP = PermMatrix(
        [1,3,2,4], [1,1.0im,1.0im,1])

# FSim is already defined in YaoExtensions
julia> @const_gate MyFSim = mat((ISWAP*
        control(2, 2, 1=>shift(-π/6)))')

julia> put(10, (4,2)=>MyFSim)
nqubits: 10
put on (4, 2)
└─ MyFSim gate
\end{lstlisting}
\end{minipage}

The macro \texttt{@const\_gate} defines a primitive gate that subtypes \texttt{ConstantGate} abstract type. It generates global gate instances \texttt{ISWAP} and \texttt{MyFSim} as well as new gate types \texttt{MyFSimGate} and \texttt{ISWAPGate} for dispatch.
This macro also binds operators properties such as \texttt{ishermitian}, \texttt{isreflexive} and \texttt{isunitary} by inspecting the given matrix representation.

In \App{app-qftblock}, we provide a more sophisticated example of extending QBIR.

\subsection{Extending the Quantum Register}

A new register type can be defined by dispatching the "instruction set" interfaces introduced in Sec.~\ref{sec:instruct}. For example, in the CUDA backend \texttt{CuYao}~\cite{cuyao} we dispatch \texttt{instruct!} and \texttt{measure!} to the \texttt{ArrayReg\{B,T,<:CuArray\}} type. Here besides the batch number \texttt{B} and data type \texttt{T}, the third type parameter \texttt{<:CuArray} specifies the storage type. The dispatch directs the instructions to the CUDA kernels written in \texttt{CUDAnative}~\cite{besard2018effective}, which significantly boosts the performance by parallelizing both for the batch and the Hilbert space dimensions and. A comparison between the batched or single
register of the parameterized circuit is shown in Sec \ref{sec:performance}. We provide more detailed examples in the developer's guide of \App{app:reading}.

\section{Applications}\label{sec:application}
The \texttt{Yao} framework has been employed in practical research projects during its development and has evolved with the requirements of research.

\texttt{Yao}  simplifies the original implementation of quantum circuit Born machine~\cite{liu2018differentiable} originally written in \texttt{ProjectQ}~\cite{steiger2016projectq} from
about 200 lines of code to less than 50 lines of code with about 1000x performance improvement 
as shown in our benchmark \Fig{fig:pcircuit-benchmark}. This simplification enabled further exploration of the algorithm in
~\cite{zeng2019learning} with the much simpler codebase.

The tensor network inspired quantum circuits described in~\cite{huggins2018towards,liu2019variational} allow the study of large systems with a reduced number of qubits. For example, one can solve the ground state of a $6\times 6$ frustrated Heisenberg lattice model with only 12 qubits. These circuits can also compress a quantum state to the hardware with fewer qubits~\cite{wwspaper}. These applications need to measure and reuse qubits in the circuits. Thus, one can not take \texttt{nshots} measurements to the wavefunction. Instead, one constructs a batched quantum register with \texttt{nbatch} states and samples bitstrings in parallel. \texttt{Yao}'s SPMD friendly design and CUDA backend significantly simplifies the implementation and boosts performance.

Automatic differentiation can also be used for gate learning. It is well-known that a general two-qubit gate can be compiled to a fixed gate instruction set that includes single-qubit gates and CNOT gates~\cite{Shende2004}.
Given a circuit structure, one can approximate an arbitrary U(4) unitary (up to a global phase) instantly by gradient optimization of the operator fidelity. The code can be found in \App{app:gatelearning}.

A recent project extending VQE~\cite{Peruzzo2014} to thermal quantum states~\cite{betaVQE} integrates \texttt{Yao}  seamlessly with the differentiable programming package \texttt{Zygote}~\cite{zygote2019mike}. \texttt{Yao}'s efficient AD engine and batched quantum register support allow joint training of quantum circuit and classical neural network effortlessly. 

\section{Roadmap} \label{sec:roadmap}

\subsection{Hardware Control}
Hardware control is one of \texttt{Yao}'s next major missions. In principle, \texttt{Yao} already has a lightweight tool \texttt{YaoScript} (\App{app:yaoscript}) to serialize QBIR to the data stream for dump/load and internet communication, which can be used to control devices on the cloud. 

For even better integration with existing protocols, we plan to support the parsing and code generation for other low level languages
such as OpenQASM~\cite{cross2017open}, eQASM~\cite{fu2019eqasm} and Quil~\cite{smith2016practical}. As an ongoing project towards this end, we and the author of the general-purpose parser generator \texttt{RBNF}~\cite{rbnf}
developed a QASM parser and codegen in \texttt{YaoQASM}~\cite{yao2019qasm}. It allows one to parse the OpenQASM~\cite{cross2017open} to QBIR and dump QBIR to OpenQASM, which can be used for controlling devices in the IBM Q Experience~\cite{steiger2016projectq, Garcia2019}. 

\subsection{Circuit Compilation and Optimization}
Circuit compilation and optimization is a key topic for quantum computing towards practical experiments. A new language interface along with a compiler for \texttt{Yao}~\cite{yao2020ir} is under development. It should be more compilation friendly with a design based on Julia's native abstract syntax tree. By integrating with Julia compiler, it will allow Yao to model quantum channels in a seamlessly way.

On the other hand, quantum circuit simplification and optimization are crucial for reducing the cost of both simulations and experiments. The ongoing circuit simplification project~\cite{yao2020zx} will also support better pattern matching and term rewriting system with support of ZX calculus~\cite{kissinger2020Pyzx}. This will allows smarter and more systematic circuit simplifications such as the ones in Refs.~\cite{iten2019efficient,maslov2008quantum,kissinger2019tcount}. 

\subsection{Noisy Simulation}
Noise important for simulation of near-term quantum devices.
Currently, \texttt{Yao} does not support noisy simulations directly. However,
a batched register in \texttt{Yao} be conveniently converted to a reduced density matrix.
By porting \texttt{Yao} with \texttt{QuantumInformation.jl}~\cite{Gawron2018}, one can carry out noisy simulation with density matrices. One can find a blog in \App{app:reading}.

\subsection{Tensor Networks Backend}
Quantum circuits are special cases of tensor networks with the unitary constraints on the gates.
Thus, tensor network algorithms developed in the quantum many-body community are potentially useful
for simulating quantum circuits, especially the shallow ones with a large number of qubits where the state vector does not fit into memory~\cite{boixo2017simulation,chen2018classical,guo2019general}. In this sense, one can perform more efficient simulations at a larger scale by exploring low-rank structures in the tensor networks with a trade-off of almost negligible errors~\cite{pan2019contracting}. 

Besides serving as an efficient simulation backend, the tensor networks are also helpful for quantum machine learning, given the fact that they have already found many applications in various machine learning tasks~\cite{Stoudenmire2016, han2018unsupervised, PhysRevB.99.155131, glasser2019expressive, bradley2019modeling}. As an example, one can envision, training a unitary tensor network with classical algorithms and then load it to an actual quantum device for fast sampling. In this regard, quantum devices become specialized inference hardware. 

The ongoing project \texttt{YaoTensorNetwork}~\cite{yaotn} provides utilities to dump a quantum circuit into a tensor network format. There are already working examples of generating tensor networks for QFT, variational circuits~\cite{Kandala2017}, and for demonstrating the quantum supremacy on random circuits~\cite{boixo2018characterizing, Google2019}. The dumped tensor networks can be further used for quantum circuit simplification~\cite{Backens2014} and quantum circuit simulation based on exact or approximated tensor network contractions.

\section{Summary}\label{sec:sum}
We have introduced \texttt{Yao}, an open source Julia package for quantum algorithm design. \texttt{Yao} features
\begin{itemize}
    \item differentiable programming quantum circuits with a built-in AD engine leveraging reversible computing, 
    \item batched quantum registers with CUDA parallelization,
    \item symbolic manipulation of quantum circuits,
    \item strong extensibility, 
    \item top performance for relevant applications.
\end{itemize}
The quantum block abstraction of the quantum circuit is central to these features. Generic programming, which in Julia is based on the type system and multiple dispatch, is key to the extensibility and efficiency of \texttt{Yao}. Along with the roadmap Sec.~\ref{sec:roadmap}, \texttt{Yao} is evolving towards an even more versatile framework for quantum computing research and development.

\section{Acknowledgement}
Thanks to Jin Zhu for the Chinese calligraphy of \texttt{Yao}'s logo. The QASM compilation was greatly aided by
Taine Zhao's effort on the Julia parser generator \texttt{RBNF.jl}, we appreciate
his help and insightful discussion about compilation. This work owes much to enlightening conversation
and help from open source community including: Tim Besard and Valentin Churavy for their work on the CUDA
transpiler \texttt{CUDAnative} and suggestions on our CUDA backend implementation, Mike Innes and Harrison Grodin for their
helpful discussion about automatic differentiation and symbolic manipulation, Juan Gomez, Christopher J. Wood, Damian Steiger, Damian Steiger, Craig Gidney, corryvrequan, Johannes Jakob Meyer and Nathan Killoran for reviewing the performance benchmarks~\cite{quantum2019benchmark}.
We thank Divyanshu Gupta for integrating \texttt{Yao} with \texttt{DifferentialEquations.jl}~\cite{diffeq.jl}, Wei-Shi Wang, Yi-Hong Zhang, Tong Liu, Yu-Kun Zhang, and Si-Rui Lu for being beta users and offering valuable suggestions. We thank Roger Melko for helpful suggestions of this manuscript, Hao Xie and Arthur Pesah for proofreading this manuscript. 
The first author would like to thank Roger Melko, Miles Stoudenmire, Xi Xiong for their
kindly help on his Ph.D. visa issue during the development.
\texttt{Yao}'s development is supported by the National Natural Science Foundation of China under
the Grant No.~11774398, No.~11747601 and No.~11975294, the Ministry of Science and Technology of China under the Grant
No. 2016YFA0300603 and No. 2016YFA0302400, the Strategic Priority Research Program of Chinese Academy of Sciences Grant No. XDB28000000, and Huawei Technologies under Grant No. YBN2018095185.

\Urlmuskip=0mu plus 1mu\relax
\bibliographystyle{unsrtnat}
\bibliography{ref}

\onecolumn\newpage
\appendix

\section{Using external libraries}\label{app:external}

It is straightforward to make use of external libraries written in other languages. 
For example, the following codes import modules from the \texttt{OpenFermion}~\cite{Mcclean2017} and construct a molecule Hamiltonian in Julia. 

\begin{lstlisting}
using PyCall

of_hamil = pyimport("openfermion.hamiltonians")
of_trsfm = pyimport("openfermion.transforms")
of_pyscf = pyimport("openfermionpyscf")

diatomic_bond_length = 1.0
geometry = [("H", (0., 0., 0.)), ("H", (0., 0., diatomic_bond_length))]
basis = "sto-3g"
multiplicity = 1
charge = 0
description = string(diatomic_bond_length)

molecule = of_hamil.MolecularData(geometry, basis, multiplicity, charge, description)
molecule = of_pyscf.run_pyscf(molecule,run_scf=1,run_fci=1)

m_h = molecule.get_molecular_hamiltonian()
nbits = m_h.n_qubits
jw_h = of_trsfm.jordan_wigner(of_trsfm.get_fermion_operator(m_h))
\end{lstlisting}

Up to here, the codes follow the \texttt{OpenFermion}'s tutorial closely. Next, we construct a quantum block representation of the molecule Hamiltonian. One can then use the Hamiltonian block as wish in \texttt{Yao}, such as exact diagonalization (Listing~\ref{lst:heisenberg}), time evolution (Listing~\ref{lst:cache-te}) or VQE (Listing \ref{lst:ad}). 

\begin{lstlisting}
julia> using Yao 
julia> function yao_hamiltonian(nbits, jw_h)
          gates = Dict("X"=>X, "Y"=>Y, "Z"=>Z)
          h = Add{nbits}()
          for (k, v) in jw_h.terms
             push!(h, v*kron(nbits, [site+1 => gates[opname] for (site, opname) in k]...))
          end
          return h
        end
        
julia> yao_hamiltonian(nbits, jw_h)
+
├─ [scale: 0.10622904490856075 + 0.0im] kron
│     ├─ 2=>Z
│     └─ 4=>Z
├─ [scale: -0.04919764587136755 + 0.0im] kron
│     ├─ 1=>X
│     ├─ 2=>X
│     ├─ 3=>Y
│     └─ 4=>Y
⋅
⋅
⋅
└─ [scale: 0.13716572937099503 + 0.0im] kron
      └─ 2=>Z
\end{lstlisting}

\newpage
\section{Builtin Block Types}\label{app:block}
One can inspect the builtin block types in \texttt{Yao} with the following codes.
\begin{lstlisting}
julia> using Yao, AbstractTrees

julia> AbstractTrees.children(x::Type) = subtypes(x)

julia> AbstractTrees.print_tree(AbstractBlock)
AbstractBlock
├─ CompositeBlock
│  ├─ AbstractContainer
│  │  ├─ ControlBlock
│  │  ├─ PutBlock
│  │  ├─ RepeatedBlock
│  │  ├─ Subroutine
│  │  └─ TagBlock{BT,N} where N where BT<:AbstractBlock
│  │     ├─ CachedBlock{ST,BT,N} where N where BT<:AbstractBlock where ST
│  │     ├─ Daggered
│  │     ├─ Scale
│  │     └─ NoParams
│  ├─ Add
│  ├─ ChainBlock
│  ├─ KronBlock
│  └─ UnitaryChannel
└─ PrimitiveBlock
   ├─ ConstantGate
   │  ├─ HGate
   │  ├─ I2Gate
   │  ├─ SWAPGate
   │  ├─ TGate
   │  ├─ XGate
   │  ├─ YGate
   │  ├─ CNOTGate
   │  ├─ CZGate
   │  ├─ IGate
   │  ├─ P0Gate
   │  ├─ P1Gate
   │  ├─ PdGate
   │  ├─ PuGate
   │  ├─ SGate
   │  ├─ SdagGate
   │  ├─ TdagGate
   │  ├─ ToffoliGate
   │  └─ ZGate
   ├─ GeneralMatrixBlock
   ├─ Measure
   ├─ PhaseGate
   ├─ RotationGate
   ├─ ShiftGate
   ├─ TimeEvolution
   └─ TrivialGate
      └─ IdentityGate
\end{lstlisting}

More blocks, such as \texttt{SqrtX} and \texttt{FSimGate}, are avaiblable in \texttt{YaoExtensions}.

\newpage
\section{Symbolic Computation: Shor's 9 qubit code}\label{app:symbolic}

The well-known Shor's 9 qubit code can correct any single-qubit error $E$ and restores the state $|\psi\rangle$.

$$
\Qcircuit @C=1.2em @R=.1em {
   & |\psi\rangle & & \ctrl{3} & \ctrl{6} & \gate{H} & \ctrl{1} & \ctrl{2} & \multigate{8}{E} & \ctrl{1} & \ctrl{2} & \targ     & \gate{H} & \ctrl{3} & \ctrl{6} & \targ     & \qw & |\psi\rangle\\
   & |0\rangle & & \qw      & \qw      & \qw      & \targ    & \qw      & \ghost{E}        & \targ    & \qw      & \ctrl{-1} & \qw      & \qw      & \qw      & \qw       & \qw \\
   & |0\rangle & & \qw      & \qw      & \qw      & \qw      & \targ    & \ghost{E}        & \qw      & \targ    & \ctrl{-2} & \qw      & \qw      & \qw      & \qw       & \qw \\
   & |0\rangle & & \targ    & \qw      & \gate{H} & \ctrl{1} & \ctrl{2} & \ghost{E}        & \ctrl{1} & \ctrl{2} & \targ & \gate{H}     & \targ    & \qw      & \ctrl{-3} & \qw \\
   & |0\rangle & & \qw      & \qw      & \qw      & \targ    & \qw      & \ghost{E}        & \targ    & \qw      & \ctrl{-1} & \qw      & \qw      & \qw      & \qw       & \qw \\
   & |0\rangle & & \qw      & \qw      & \qw      & \qw      & \targ    & \ghost{E}        & \qw      & \targ    & \ctrl{-2} & \qw      & \qw      & \qw      & \qw       & \qw \\
   & |0\rangle & & \qw      & \targ    & \gate{H} & \ctrl{1} & \ctrl{2} & \ghost{E}        & \ctrl{1} & \ctrl{2} & \targ & \gate{H}     & \qw      & \targ    & \ctrl{-6} & \qw \\
   & |0\rangle & & \qw      & \qw      & \qw      & \targ    & \qw      & \ghost{E}        & \targ    & \qw      & \ctrl{-1} & \qw      & \qw      & \qw      & \qw       & \qw\\
   & |0\rangle & & \qw      & \qw      & \qw      & \qw      & \targ    & \ghost{E}        & \qw      & \targ    & \ctrl{-2} & \qw      & \qw      & \qw      & \qw       & \qw\\
}
$$

The circuit can be constructed by the following code.

\begin{lstlisting}
using Yao

shor(E) = chain(9,
    # encode circuit
    cnot(1, 4), cnot(1, 7),
    put(1=>H), put(4=>H), put(7=>H),
    cnot(1,2), cnot(1,3), cnot(4,5), cnot(4,6), cnot(7,8), cnot(7,9),

    E, # the error

    # decode circuit
    cnot(1,2), cnot(1,3), cnot((2, 3), 1),
    cnot(4,5), cnot(4,6), cnot((5, 6), 4),
    cnot(7,8), cnot(7,9), cnot((8, 9), 7),

    put(1=>H), put(4=>H), put(7=>H),
    cnot(1, 4), cnot(1, 7), cnot((4, 7), 1)
    )
\end{lstlisting}

Now we can check whether it can correct a given error by doing symbolic computation on
an arbitrary 1-qubit pure quantum state $\alpha|0\rangle + \beta |1\rangle$ and a specific weight-9 error.

\begin{lstlisting}
julia> using Yao, SymEngine

julia> @vars α β
(α, β)

julia> s = α * ket"0" + β * ket"1" |> addbits!(8)
α|000000000> + β|000000001>

julia> E = kron(1=>X, 2=>Z, 3=>Z, 4=>X, 5=>Z, 6=>Z, 7=>X, 8=>Z, 9=>Z);

julia> s |> shor(E) |> partial_tr(2:9) |> expand
α|0> + β|1>
\end{lstlisting}

Yeah! 

\newpage

\section{Yao Script}\label{app:yaoscript}
We introduce \texttt{YaoScript} as an alternating way to define quantum circuits.
We provide a Julia non-standard string literal \texttt{@yao\_str} for circuit parsing, which can be invoked through \texttt{yao"..."}.
This macro reads and parses a string to a QBIR. With \texttt{YaoScript}, one can dump (load) a circuit to (from) a file or internet data stream. The \texttt{YaoScript} appears like native Julia code but will be parsed by the macro to prevent code injection.

The error correction circuit used in \App{app:symbolic} can be defined as
\begin{lstlisting}
using Yao
c = yao"""let nqubits=9, version="0.6.0"
    begin # encode circuit
        1=>C, 4=>X
        1=>C, 7=>X
        1=>H, 4=>H, 7=>H
        1=>C, 2=>X
        1=>C, 3=>X
        4=>C, 5=>X
        4=>C, 6=>X
        7=>C, 8=>X
        7=>C, 9=>X
    end

    # the error
    1=>X, 2=>Z, 3=>Z, 4=>X, 5=>Z, 6=>Z, 7=>X, 8=>Z, 9=>Z

    begin # decode circuit
        1=>C, 2=>X
        1=>C, 3=>X
        2=>C, 3=>C, 1=>X
        4=>C, 5=>X
        4=>C, 6=>X
        5=>C, 6=>C, 4=>X
        7=>C, 8=>X
        7=>C, 9=>X
        8=>C, 9=>C, 7=>X

        1=>H, 4=>H, 7=>H
        1=>C, 4=>X
        1=>C, 7=>X
        4=>C, 7=>C, 1=>X
    end
end"""
\end{lstlisting}
In this script, it is easy to find the following correspondance with QBIR
\begin{table}[h!]\centering
\begin{minipage}{\columnwidth}
\ra{1.3}
    \scalebox{1.1}{
        \begin{tabularx}{0.9\textwidth}{X X}\toprule
            \textbf{Script} & \textbf{Block}\\
            \hline
            1=>C, 2=>X & \texttt{control(9, 1, 2=>X)}\\
            1=>H, 4=>H, 7=>H & \texttt{kron(9, 1=>H, 4=>H, 7=>H)}\\
            1=>H & \texttt{put(9, 1=>H)}\\
            beigin ... end & \texttt{chain(9, [...])}\\
            \bottomrule
        \end{tabularx}
    }
    \caption{The correspondence between \texttt{YaoScript} and QBIR.}\label{tbl-mattype}
\end{minipage}
\end{table}

\newpage

\section{Matrix Types in Yao}\label{app:sparse}
\Tbl{tbl-mattype}  summaries the matrix types used for the basic quantum gates.

\begin{table}[h!]\centering
\begin{minipage}{\columnwidth}
\ra{1.3}
    \scalebox{1.1}{
        \begin{tabularx}{0.9\textwidth}{X X}\toprule
            \textbf{Gate} & \textbf{Matrix Type}\\
            \hline
            I2 & \texttt{IMatrix}\\
            Z, T, S, Rz & \texttt{Diagonal}\\
            X, Y, CNOT, CZ, SWAP & \texttt{PermMatrix}\\
            P0, P1, Pu, Pd, PSwap & \texttt{SparseMatrixCSC}\\
            H, Rx, Ry & \texttt{Matrix}\\
            \bottomrule
        \end{tabularx}
    }
    \caption{Matrix types of gates in \texttt{Yao}.}\label{tbl-mattype}
\end{minipage}
\end{table}
The \texttt{SparseMatrixCSC} type is provided in Julia's builtin~\texttt{SparseArrays}. The identity matrix \texttt{IMatrix} and general permutation matrix \texttt{PermMatrix}~\cite{generalperm} are defined in \texttt{LuxurySparse.jl}~\cite{luxurysparse}. The \texttt{PermMatrix} allows having values other than one in the non-zero entries.
For example, the matrix of ISWAP gate defined in Sec.~\ref{sec:extend-ir-nodes} of the main text
\begin{lstlisting}
julia> PermMatrix([1,3,2,4], [1,1.0im,1.0im,1])
4×4 PermMatrix{Complex{Float64},Int64,Array{Complex{Float64},1},Array{Int64,1}}:
 1.0+0.0im     0          0          0     
    0          0       0.0+1.0im     0     
    0       0.0+1.0im     0          0     
    0          0          0       1.0+0.0im
\end{lstlisting}
where the first argument represents the column indices and the second argument the entries.

These type specifications for quantum gates allow fast arithmetics.  \Tbl{tbl-mulkron} lists the type conversion under matrix multiplication, Kronecker product, and addition operations.

\begin{table}[h!]\centering
\begin{minipage}{\columnwidth}
\ra{1.3}
    \scalebox{1.1}{
        \begin{tabularx}{0.9\textwidth}{X X X X X X}\toprule
              & I   & D   & P   & S   & M  \\
            I & I/I/D/I & D/D/D/D & P/P/S/D & S/S/S/D & M/S/M/D\\
            D & D/D/D/D & D/D/D/D & P/P/S/D & S/S/S/D & M/S/M/D\\
            P & P/P/S/D & P/P/S/D & P/P/S/P & S/S/S/P & M/S/M/P\\
            S & S/S/S/D & S/S/S/D & S/S/S/P & S/S/S/S & M/S/M/S\\
            M & M/S/M/D & M/S/M/D & M/S/M/P & M/S/M/S & M/S/M/M\\
            \bottomrule
        \end{tabularx}
    }
    \caption{Matrix types conversion under matrix multiplication (*)/kronecker product (kron)/addition (+)/hadamard product (\texttt{.*}). Here I, D, P, S, M stands for \texttt{IMatrix}, \texttt{Diagonal}, \texttt{PermMatrix}, \texttt{SpasreMatrixCSC} and \texttt{Matrix} respectively.}\label{tbl-mulkron}
\end{minipage}
\end{table}

Besides these specialised sparse matrices, \texttt{Yao.AD} uses low rank matrix types for backpropagation, c.f. \Eq{eq:outer-product} in the main texts. For this we define the \texttt{OuterProduct} matrix type for both memory and computation efficiency.

\newpage
\section{Extending QBIR: emulating QFT circuit as an example}\label{app-qftblock}

In most cases, one can build quantum circuits using basic blocks in \texttt{YaoBlocks} such as \texttt{put}, \texttt{control} in the Listing~\ref{lst:qft}. In certain cases, one needs to define a new QBIR node, e.g., to
dispatch a more specialized simulation method. For example, one can emulate QFT by using highly optimized FFT on classical computers~\cite{haner2016high}. Since the nodes are just normal Julia type, it is straightforward to do this in \texttt{Yao}. One can define a new block type by subtyping from the primitive block.
In principle, the only thing to make it work is to define its matrix representation.

\begin{lstlisting}
using Yao

struct QFT{N} <: PrimitiveBlock{N} end
QFT(n::Int) = QFT{n}()
Yao.mat(::Type{T}, x::QFT) where T = mat(T, qft(nqubits(x)))
\end{lstlisting}

where \texttt{qft(n)} can be the function defined in Listing~\ref{lst:qft}. In this way, one can wrap an implementation of the QFT as a primitive block.

However, to emulate QFT in a more efficient way, we can overload the \texttt{apply!} method
using the classical inverse FFT for the QFT circuit.
\begin{lstlisting}
using FFTW, LinearAlgebra, BitBasis

function Yao.apply!(r::ArrayReg, x::QFT)
    α = sqrt(length(statevec(r)))
    invorder!(r)
    lmul!(α, ifft!(statevec(r)))
    return r
end
\end{lstlisting}

With this minimal definition, \texttt{Yao} is able to make use of builtin methods
to infer its properties such as \texttt{ishermitian}, \texttt{isunitary}, \texttt{isreflexive} and \texttt{iscommute}
and other functionalities. For example, the inverse QFT can be simply obtained
by \texttt{QFT(4)\textquotesingle}. After the extension, the \texttt{QFT} type will appear in the type tree shown in \App{app:block} under the \texttt{PrimitiveBlock}. 

Both the faithful simulation approach introduced in the main text and this classical emulation have been
included in \textbf{YaoExtensions} as \texttt{qft\_circuit} (for faithful classical simulation) and \texttt{QFT}
(for FFT emulation).

\newpage
\section{Symbolic Differentiation}\label{app:symad}

The following program computes the gradient of a Heisenberg Hamiltonian with respect to circuit parameters symbolically and analytically with \texttt{Yao}'s AD engine. 

\begin{lstlisting}
julia> using Yao, YaoExtensions, SymEngine

julia> @vars α β γ
(α, β, γ)

julia> circuit = chain(put(3, 2=>Rx(α)), control(3, 2, 1=>Ry(β)), put(3, (1,2)=>rot(kron(X, X), γ)))
nqubits: 3
chain
├─ put on (2)
│  └─ rot(XGate, α)
├─ control(2)
│  └─ (1,) rot(YGate, β)
└─ put on (1, 2)
   └─ rot(KronBlock{2,XGate}, γ)

julia> h = heisenberg(3);

julia> energy_symbolic = expect(h, zero_state(Basic, 3)=>circuit)
-(-I*sin((1/2)*γ)*cos((1/2)*α) - ... - sin((1/2)*γ)*sin((1/2)*β)*sin((1/2)*α))^2

julia> grad_symbolic = expect'(h, zero_state(Basic, 3)=>circuit).second
3-element Array{Basic,1}:
 -((sin((1/2)*β)*(-sin((1/2)*γ)* ... + cos((1/2)*γ)^2*cos((1/2)*β)*sin((1/2)*α))))
 2*((-1/2)*sin((1/2)*β)*(I*sin((1/2)*γ)* ... + I*cos((1/2)*γ)*sin((1/2)*β)*sin((1/2)*α)))))
 -((cos((1/2)*γ)*cos((1/2)*α) - ... - 3*sin((1/2)*γ)*sin((1/2)*β)*sin((1/2)*α)))

julia> assign = Dict(α=>0.5, β=>0.7, γ=>0.8);

julia> energy_eval = subs(energy_symbolic, assign...)
1.9542144196548 + -0.0*I

julia> grad_eval = map(x->subs(x, assign...), grad_symbolic)
3-element Array{Basic,1}:
          -1.22808300500511
 -0.311108582564352 + 0.0*I
          -1.56563863069374

julia> circuit_numeric = subs(Float64, circuit, assign...)
nqubits: 3
chain
├─ put on (2)
│  └─ rot(X gate, 0.5)
├─ control(2)
│  └─ (1,) rot(Y gate, 0.7)
└─ put on (1, 2)
   └─ rot(KronBlock{2,XGate}, 0.8)

julia> energy_numeric = expect(h, zero_state(3)=>circuit_numeric)
1.9542144196547988 + 0.0im

julia> grad_numeric = expect'(h, zero_state(3)=>circuit_numeric).second
3-element Array{Float64,1}:
 -1.2280830050051128 
 -0.31110858256435187
 -1.5656386306937393 

\end{lstlisting}

\newpage
\section{Gate Learning}\label{app:gatelearning}
Given a target unitary matrix and a parameterized circuit, we can learn the gate parameters by minimizing the distance between the two unitaries. Gate learning is useful for quantum compiling, diagonalization~\cite{LaRose2019, Cirstoiu2019}, and automated quantum algorithm design~\cite{Cincio2018}.

\begin{lstlisting}
julia> using YaoExtensions, Yao, Optim

julia> function learn_u4(u::AbstractBlock; niter=100)
           ansatz = general_U4()
           params = parameters(ansatz)
           println("initial fidelity = $(operator_fidelity(u,ansatz))")
           Optim.optimize(x->-operator_fidelity(u, dispatch!(ansatz, x)),
                   (G, x) -> (G .= -operator_fidelity'(u, dispatch!(ansatz, x))[2]),
                   parameters(ansatz), Optim.LBFGS(), Optim.Options(iterations=niter))
           println("final fidelity = $(operator_fidelity(u,ansatz))")
           return ansatz
       end;

julia> u = matblock(rand_unitary(4));

julia> c = learn_u4(u; niter=150)
initial fidelity = 0.1387857997337068
final fidelity = 0.9999999999999997
nqubits: 2
chain
├─ put on (1)
│  └─ chain
│     ├─ rot(Z, 0.19669849872229161)
│     ├─ rot(Y, 1.7777036486174476)
│     └─ rot(Z, 0.5789834641533261)
├─ put on (2)
│  └─ chain
│     ├─ rot(Z, 1.5080795415675794)
│     ├─ rot(Y, 0.75809724318303)
│     └─ rot(Z, -2.131912688745343)
⋅
⋅
⋅
├─ put on (1)
│  └─ chain
│     ├─ rot(Z, 2.179888127260871)
│     ├─ rot(Y, -0.348765381903638)
│     └─ rot(Z, -1.4060784496633136)
└─ put on (2)
   └─ chain
      ├─ rot(Z, 0.40168928969314993)
      ├─ rot(Y, 0.8982267608669345)
      └─ rot(Z, -0.9468001921281308)
\end{lstlisting}
In this example, the loss is the operator fidelity defined as $F(U_1, U_2)=|{\rm Tr} (U_1^\dagger U_2)|/d$~\cite{Wang2009}, with $d$ the size of Hilbert space.
\texttt{operator\_fidelity\textquotesingle(u1, u2)} returns the gradients of parameters in \texttt{u1} and \texttt{u2}.
We use the LBFGS optimizer~\cite{Byrd1995} provided in \texttt{Optim.jl}~\cite{mogensen2018optim} for optimization. The circuit ansatz (\texttt{general\_U4}) is the minimal universal two-qubit gate decomposition of Ref.~\cite{Shende2004} defined in \texttt{YaoExtensions}. 

\newpage
\section{Further Readings}\label{app:reading}

\subsection{Quantum Algorithms}
A growing list of quantum algorithms that are implemented in \texttt{Yao} is shown below.

\begin{itemize}
    \item \href{https://github.com/QuantumBFS/YaoExtensions.jl/blob/v0.2.0/src/easybuild/qft_circuit.jl}{Quantum Fourier transformation}
    \item \href{https://github.com/QuantumBFS/QuAlgorithmZoo.jl/blob/v0.1.0/src/PhaseEstimation.jl}{Phase estimation}
    \item \href{https://github.com/QuantumBFS/QuAlgorithmZoo.jl/blob/v0.1.0/examples/Shor}{Shor's algorithm}
    \item \href{https://github.com/QuantumBFS/QuAlgorithmZoo.jl/blob/v0.1.0/src/hamiltonian_solvers.jl}{Imaginary time evolution quantum eigensolver}
    \item \href{https://github.com/QuantumBFS/QuAlgorithmZoo.jl/blob/v0.1.0/examples/VQE}{Variational quantum eigensolver}
    \item \href{https://github.com/QuantumBFS/QuAlgorithmZoo.jl/blob/v0.1.0/src/HadamardTest.jl}{Hadamard test}
    \item \href{https://github.com/QuantumBFS/QuAlgorithmZoo.jl/blob/v0.1.0/examples/QSVD}{Quantum singular value decomposition}
    \item \href{https://github.com/QuantumBFS/QuAlgorithmZoo.jl/blob/v0.1.0/examples/HHL}{HHL algorithm for linear systems of equations}
    \item \href{https://github.com/QuantumBFS/QuAlgorithmZoo.jl/blob/v0.1.0/examples/QAOA}{Quantum approximate optimization algorithm}
    \item \href{https://github.com/QuantumBFS/QuAlgorithmZoo.jl/blob/v0.1.0/examples/QCBM}{Quantum circuit Born machine for generative modeling}
    \item \href{https://github.com/QuantumBFS/QuAlgorithmZoo.jl/blob/v0.1.0/examples/QuGAN}{Quantum generative adversarial circuits}
    \item \href{https://github.com/QuantumBFS/QuAlgorithmZoo.jl/blob/v0.1.0/examples/Grover}{Grover search}
    \item \href{https://github.com/QuantumBFS/QuDiffEq.jl}{Quantum ordinary differential equation}
    \item \href{https://github.com/GiggleLiu/QuantumPEPS.jl}{Qubit efficient VQE with tensor network inspired circuits}
\end{itemize}

\subsection{Developer's guide}
\texttt{Yao}'s tutorials contain several useful examples for prospective developers.

\begin{itemize}
    \item \href{http://tutorials.yaoquantum.org/v0.6/generated/developer-guide/1.extending-register/}{Extending the register type: the echo register}    
    \item \href{http://tutorials.yaoquantum.org/v0.6/generated/developer-guide/2.cuda-acceleration/}{Implementing the SWAP gate in CUDA}
\end{itemize}

\subsection{Porting Yao to other parts of Julia ecosystem}\label{app:yaoeco}
\begin{itemize}
    \item \href{https://github.com/QuantumBFS/QuAlgorithmZoo.jl/blob/v0.1.0/examples/PortZygote/gate\_learning.jl}{Porting the builtin AD engine to Zygote for  gate learning}.
    \item \href{https://github.com/QuantumBFS/QuAlgorithmZoo.jl/blob/v0.1.0/examples/PortQuantumInfromation}{Porting Yao to QuantumInformation~\cite{Gawron2018}}
\end{itemize}

\end{document}